\documentclass{emulateapj}

\usepackage{natbib}
\usepackage{graphicx}

\newcommand{\tnm}{\tablenotemark}
\newcommand{\tnt}{\tablenotetext}

\newcommand{\ergs}{ergs s$^{-1}$}
\newcommand{\flux}{ergs cm$^{-2}$ s$^{-1}$}
\newcommand{\pers}{s$^{-1}$}
\newcommand{\cdens}{cm$^{-2}$}
\newcommand{\vdens}{cm$^{-3}$}

\newcommand{\xmm}{{\it XMM}}
\newcommand{\xmmnewton}{{\it XMM-Newton}}
\newcommand{\chandra}{{\it Chandra}}
\newcommand{\asca}{{\it ASCA}}
\newcommand{\rxte}{{\it RXTE}}
\newcommand{\rosat}{{\it ROSAT}}
\newcommand{\bepposax}{{\it BeppoSAX}}
\newcommand{\ginga}{{\it Ginga}}
\newcommand{\exosat}{{\it EXOSAT}}
\newcommand{\einstein}{{\it Einstein}}

\newcommand{\mekal}{{\it MEKAL}}

\newcommand{\rxj}{RX J0059.2--7138}
\newcommand{\xtej}{XTE J0111.2--7317}
\newcommand{\fouru}{4U 1626--67}
\newcommand{\aofive}{A 0538--66}
\newcommand{\exo}{EXO 053109--6609.2}

\newcommand{\phip}{$\phi_{35}$}

\newcommand{\im}{\item}

\newcommand{\alfven}{$\rm{Alfv\grave{e}n}$}

\newcommand{\dgr}{$^{\circ}$}

\newcommand{\msun}{M_{\sun}}

\newcommand{\W}{\hphantom{0}}
\newcommand{\Wp}{\hphantom{.}}

\begin{document}
\slugcomment{Accepted for publication in the Astrophysical Journal}

\title{Origin of the soft excess in X-ray pulsars}
\shorttitle{SOFT EXCESS IN X-RAY PULSARS}
\shortauthors{HICKOX, NARAYAN, \& KALLMAN}
\author{Ryan C. Hickox}
\author{Ramesh Narayan}
\affil{Harvard-Smithsonian Center for Astrophysics, 60 Garden Street,
 Cambridge, MA 02138}
\and
\author{Timothy R. Kallman}
\affil{NASA Goddard Space Flight Center, Laboratory for High Energy
Astrophysics, Code 665, Greenbelt, MD 20771}

\begin{abstract}
The spectra of many X-ray pulsars show, in addition to a power law, a
low-energy component that has often been modeled as a blackbody with
$kT_{\rm BB} \sim$ 0.1 keV.  However the physical origin of this soft
excess has remained a mystery.  We examine a sample of well-studied,
bright X-ray pulsars, which have been observed using \rosat, \asca,
\ginga, \rxte, \bepposax, \chandra, and \xmmnewton.  In particular, we
consider the Magellanic Cloud pulsars SMC X-1, LMC X-4, \xtej, and
\rxj\ and the Galactic sources Her X-1, \fouru, Cen X-3, and Vela
X-1. We show that the soft excess is a very common if not ubiquitous
feature intrinsic to X-ray pulsars.  We evaluate several possible
mechanisms for the soft emission, using theoretical arguments as well
as observational clues such as spectral shapes, eclipses, pulsations
of the soft component, and superorbital modulation of the source flux.
We find that reprocessing of hard X-rays from the neutron star by the
inner region of the accretion disk is the only process that can
explain the soft excess in all the pulsars with $L_{\rm X} \gtrsim
10^{38}$ \ergs.  Other mechanisms, such as emission from diffuse gas
in the system, are important in less luminous objects.
\end{abstract}

\section{Introduction}
X-ray binary pulsars (XBPs) have provided a fruitful laboratory for
high-energy astrophysics since their discovery in the 1970's
\citep[for a review see][]{naga89}.  Many XBPs have been observed
extensively, and are useful for studying a variety of astrophysical
processes, including accretion onto compact objects, binary
interactions, and the effects of strong magnetic fields.

In recent years, studies of XBPs with orbiting observatories such as
\rosat, \asca, \ginga, \rxte, \bepposax, \chandra, and \xmmnewton\
have allowed for closer scrutiny of the relevant physics.  In
particular, most bright XBPs outside the Galaxy (for which soft X-rays
are not heavily absorbed by Galactic gas) have shown a marked soft
excess in their X-ray spectra above the standard power-law model
\citep{naga02}.  Authors have proposed a variety of simple models to
describe this soft excess, but there has been considerable debate over
the origin of this feature in a number of sources.  For example, the
soft component in LMC X-4 has been modeled as a thermal
bremsstrahlung-shaped fan beam \citep{woo96}, power-law emission from
the upper accretion column \citep{paul02}, diffuse bremsstrahlung
emission, and/or reprocessing by accreting material at the
magnetosphere \citep{laba01}.  In this paper we seek to narrow
these candidate mechanisms for the soft excess, by examining the
physical processes in light of a wide range of observations.

The Galactic XBP \objectname{Hercules X-1} has a well-studied soft
excess that can be explained by reprocessing of hard X-rays by the
inner edge of the accretion disk \citep{endo00, rams02}.  At the
moment Her X-1 is the XBP for which the soft excess is best observed
and understood, and we will use it to help guide our study of other sources.

Because the Galaxy's disk contains X-ray--absorbing gas, the soft
excess is not easy to observe in most Galactic XBPs.  This emission is
best seen in the Magellenic Clouds, at high Galactic latitude where
there is not much low-energy absorption.  The bright XBPs in these
nearby galaxies are generally more luminous than those in the Milky
Way and are close enough to obtain good spectra with a high
signal-to-noise ratio.  In a survey of {\it ASCA} observations of the SMC,
\citet{yoko03} found 30 XBPs, of which three have X-ray luminosities
large enough to extract good spectra.  All three of these sources
(\objectname{SMC X-1}, \objectname{RX J0059.2--7138}, and
\objectname{XTE J0111.2--7317}) show a clear soft excess feature in
the spectrum.  The one comparably bright XBP in the LMC (LMC X-4) also
shows a soft excess.   All four of these sources have
$L_{\rm X} \sim 10^{38}$ ergs s$^{-1}$.  

\objectname{\fouru}, \objectname{Centaurus X-3}, and \objectname{Vela
X-1} in the Galaxy, and \exo\ and \aofive\ in the LMC also show soft
excess features \citep{ange95, burd00, habe94b, habe03, mavr93}.  In
addition, a recent survey of the SMC with {\it XMM} has revealed soft
components in several lower luminosity pulsars \citep{sasa03}.  In \S\
2 we summarize observational studies of the XBPs mentioned above, and
in \S\ 3 we examine how common the soft excess feature is in the total
population of XBPs and introduce possible mechanisms for
its origin.  In \S\S\ 4--7 we evaluate these possibilities in terms of
both theory and observations.  In \S\ 8 we discuss the implications of
the results and future work, and in \S\ 9 we summarize our results.

\begin{turnpage}
\begin{deluxetable*}{lcccccccc}
\tabletypesize{\tiny}
\tablecaption{Orbital and spectral parameters for XBPs with soft excesses}
\tablewidth{0pt}
\tablehead{
\colhead{}  & 
\colhead{Her X-1\tnm{a}} &
\colhead{SMC X-1\tnm{b}} &
\colhead{LMC X-4\tnm{c}} &
\colhead{\xtej\tnm{d}} &
\colhead{\rxj\tnm{e}} &
\colhead{\fouru\tnm{f}} &
\colhead{Cen X-3\tnm{g}} &
\colhead{Vela X-1\tnm{h}}}
\startdata
Location & Galaxy & SMC & LMC & SMC & SMC & Galaxy & Galaxy & Galaxy \\
Companion & A star & B0 & O7 III-IV & Be & Be & low-mass & O6--O8 & B0.5Ib
\\
Distance (kpc) & $\sim$5 & 65 & 50 & 65 & 65 & unknown & $\sim$8 & 1.9 \\
Transient? & no & no & no & yes & yes & no & no & no \\
$P_{\rm orb}$ (d) & 1.7 & 3.9 & 1.4 & \nodata & \nodata & 0.03 & 2.1 &
8.96 \\
$P_{\rm pulse}$ (s) & 1.24 & 0.7 & 13.5 & 30.95 & 2.76 & 7.7 & 4.8 &
283 \\
$P_{\rm superorbital}$ (d) & 35 & 40-60 & 30 & \nodata & \nodata
&\nodata & \nodata & \nodata  \\
$L_{\rm X}$ (ergs s$^{-1}$) & $2\times10^{37}$ &
$2.4\times10^{38}$ & $1.2\times10^{38}$ & 
$1.8\times10^{38}$ & $2.6\times10^{38}$ & $7.7\times10^{34}D_{\rm
kpc}^2$ & $\sim$$10^{38}$ & $\sim$$10^{36}$ \\
$L_{\rm soft}/L_{\rm X}$ & 0.04 & 0.036 & 0.064 & $\sim$0.10 &
$\sim$0.35 & $\sim$0.03 & $\sim$0.5 & $\sim$0.01 \\
Soft pulses? & yes & yes & yes & yes & no & no & yes & no \\
\\
$N_{\rm H}$ ($10^{21}$ cm$^{-2}$)     & 0.05, $\sim$100\tnm{i} & 2--5 &
$\sim$0.5 & $1.8^{+0.3}_{-0.2}$    &
0.42--$0.50\pm0.05$ & $1.1\pm0.2$ & $19.5\pm0.3$ & $4.2\pm0.5$, 150\tnm{i} \\
$\Gamma_{\rm hard}$                 & 0.9--1.2 & $\sim$0.9 & 0.5--0.7
& $0.76^{+0.01}_{-0.02}$&
$0.43\pm0.05$ & $0.83\pm0.01$ & $1.208\pm0.007$ & 1.4	\\
$E_{\rm cut}$ (keV)	& 24 & $\sim$6 & 16--18 & \nodata  & $6.4^{+0.7}_{-0.9}$ &
$29\pm2$ & $13.79\pm0.13$ & \nodata \\
$E_{\rm fold}$ (keV)                    & 15 & 15--35 &18--35 & \nodata  & $9.3^{+11.2}_{-4.3}$  &
$9^{+2}_{-1}$ & $8.39\pm0.19$ &	\nodata \\
$E_{\rm Fe}$ (keV) & $\sim$6.5 & $\sim$6.5 & $\sim$6.4	& 6.4       		& \nodata  &
\nodata & $6.66\pm0.03$ & 6.4 \\
$E_{\rm cyc}$ (keV)	& $42.1\pm0.3$ & \nodata & $100^{+80}_{-15}$ & \nodata & \nodata & $33\pm1$&
$30.6\pm0.6$ & 24, 55 \\
& & & & & & & \\
Soft excess model types\tnm{j} & BB, BB+LE & BB, TB, SPL & BB, BB+TB, COM,
SPL & SPL & MEK, SPL & BB & BB & TB, lines \\
$kT_{\rm BB}$ (keV)	& 0.09--0.12 & 0.15--0.18 & 0.15 & \nodata & \nodata & \nodata &
\nodata & \nodata \\
$kT_{\rm tb/MEK/COM}$ (keV) & \nodata & $0.33\pm0.03$ (TB) &
$0.9^{+4}_{-0.3}$ (COM) & \nodata & 0.37 (MEK) &
\nodata & \nodata & \nodata \\
$kT_{\rm BB}$ plus $kT_{\rm tb}$ (keV)	& \nodata & \nodata & 0.03--0.17, 0.5--0.8 & \nodata  & \nodata & $0.30\pm0.01$ & $0.11\pm0.1$  & 0.4--$0.7\pm0.1$\\
Low-E line energies (keV) &0.95 & \nodata & 0.9, 1.9 & \nodata & \nodata & 1.05 & \nodata & \nodata
\enddata
\tnt{a}{Typical ranges of values for spectral parameters, compiled
from  fits using \bepposax\ \citep{dalf98}, \asca\
\citep{endo00}, and \xmm\ \citep{rams02}. Quoted errors for all sources are at 90\% confidence.} 
\tnt{b}{{\it ROSAT/Ginga} \citep{woo95}, \asca\ \citep{paul02}.}
\tnt{c}{{\it ROSAT/Ginga} \citep{woo96}, \bepposax\ \citep{laba01, naik04}.}
\tnt{d}{\asca\ \citep{yoko00}.}
\tnt{e}{{\it ROSAT/ASCA} \citep{kohn00}.}
\tnt{f}{\bepposax\ \citep{orla98}.}
\tnt{g}{\bepposax\ \citep{burd00}.}
\tnt{h}{\rosat\ \citep{habe94a}, \bepposax\ \citep{orla98a}, \rxte\ \citep{krey02}}
\tnt{i}{The two values for $N_{\rm
H}$ indicate interstellar absorption and partially covering
absorption instrinsic to the source.}
\tnt{j}{Spectral models used for the soft excess are: BB--blackbody,
TB--thermal bremsstrahlung, SPL--soft power law or broken power law, MEK--\mekal\ thin-thermal model,
COM--COMPST Comptonization model, LE--broad low-energy line
emission, lines--resolved individual emission lines.  Commas indicate
separate fits, plus signs indicate fits with two components.}
\end{deluxetable*}
\end{turnpage}
\newpage

\section{Observational background}
Here we review a number of observations of XBPs with a soft
excess.  Orbital and spectral parameters for each source are compiled
in Table 1.

\subsection{Her X-1}
Her X-1 was one of the first known X-ray pulsars \citep{tana72} and
is one of the few with a low-mass companion.  It has a pulse period of
1.24 s and an orbital period of 1.7 days.  The system is at high
inclination, and the A7 companion star, with a mass of $\sim$2 $\msun$,
eclipses the pulsar every orbital period \citep{midd76}.  In addition there is a 35--day
cycle in the X-ray intensity, consisting of ``main-on'' and ``short-on''
states separated by two ``low'' states in which the X-ray flux drops by
several orders of magnitude \citep{jone76}.  This modulation is
believed to be due to a tilted precessing accretion disk that
periodically blocks the central X-ray source \citep[see][for a
review]{scot99}.  Given a distance of 5 kpc, Her X-1 has $L_{\rm X} \sim
2\times 10^{37}$ \ergs.

The spectrum of Her X-1 has been extensively studied, and is generally
fitted with a cutoff power law with $\Gamma \sim 1$, a blackbody with
$kT_{\rm BB}\sim 0.1$ keV, and often broad--line emission at $\sim$1
keV.  There is also iron emission at $\sim$6.5 keV and a cyclotron
feature at $\sim$40 keV \citep{dalf98}.  The energy of the
(fundamental) cyclotron line is related to the $B$ field by
$$E_{\rm cyc} = 11.6 (B/10^{12} {\rm \ G})(1+z)^{-1} {\rm \ keV},$$
where $z$ is the gravitational redshift \citep[i.e.,][]{cobu02}.  For
surface emission from a neutron star with $R_{ns}=10$ km and $M=1.4
\msun$, we have $z\sim0.3$ \citep{cobu02}.  Usually neither $E_{\rm cyc}$ nor
$z$ is well constrained, so for simplicity we take $B \simeq
10^{12} (E_{\rm cyc}/ \rm{10 \ keV})$ G in this paper.  Thus, Her X-1
has $B\sim 4\times 10^{12}$ G.

The pulsations of the soft component are generally smoother than and
out of phase with the hard component \citep{mccr82,endo00}.  This
phase offset varies over the superorbital period \citep{rams02},
supporting the idea that the soft excess is emitted by the precessing
accretion disk. We discuss this process in detail in \S\ 7.

\subsection{The luminous Magellanic Cloud XBPs}
SMC X-1 is a luminous, eclipsing, high-mass XBP system. It was first
identified as a point source in {\it Uhuru} observations by
\citet{leon71}.  Eclipses were discovered by \citet{schr72} and X-ray
pulsations by \citet{luck76}.  The optical counterpart is Sk 160, a B0
I supergiant \citep{webs72, lill73}. The orbit is close to circular,
with orbital period $\sim$3.9 days and pulse period 0.7 s. The source
has a high $L_{\rm X} \sim 2 \times 10^{38}$ \ergs.  As in Her X-1
there are large, quasi-periodic variations of an $\sim$60 day period
between high and low states of the X-ray flux, attributed to blocking
by a precessing disk \citep{grub84}.  These variations are stable, and
the period of the cycles oscillates from $\sim$50 to $\sim$60 days
\citep{wojd98}. This oscillation is itself periodic on the order of
years \citep{clar03}. The spectrum of SMC X-1 is usually fitted as a
cutoff power law with $\Gamma \sim 0.9$, an iron line at $\sim$6.4
keV, and a soft component modeled as a broken power law,
thermal bremsstrahlung, or a blackbody with $kT_{\rm BB} \sim 0.16$
keV.  \citet{paul02} fitted four different models to the soft component
and found that all gave satisfactory fits.  Similarly to Her X-1, they
found that the soft component pulses roughly sinusoidally, out of
phase with the sharper-peaked power law.

LMC X-4 is a high-mass XBP very similar to SMC X-1, and was also
discovered by {\it Uhuru} observations \citep{giac72}.  Its companion
is an 0 III-IV star \citep{sand77}. It is eclipsing
\citep{li78,whit78} and has a pulse period of 13.5 s \citep{kell83} and
an orbital period of 1.4 days.  LMC X-4 also has a stable superorbital
period of $\sim$30 days, again attributed to a precessing disk
\citep{lang81}. About once per day the source shows flares lasting
$\sim$20--45 minutes during which the X-ray flux increases by up to
$\sim$20 times \citep[e.g.,][]{levi91}.  Like SMC X-1, the high-state
spectrum is fitted by a cutoff power law ($\Gamma \sim 0.6$), with a soft
component modeled in a variety of ways suggesting different emission
mechanisms.  In the low state, the power law hardens to $\Gamma \sim
0.1$ \citep{naik02}.  There are iron emission at $\sim 6.5$ keV and a
cyclotron feature at $\sim$100 keV \citep{laba01} giving
$B \sim 10^{13}$ G.  Similar to SMC X-1, the soft
component pulses more smoothly than the hard component and is perhaps
out of phase with it \citep{woo96, paul02}.  For both SMC X-1 and LMC
X-4, \citet{paul02} argue that the soft component is likely not
thermal in origin and is best described by power-law emission from
the upper accretion column.

\xtej\ is a bright, transient XBP in the SMC with a probable Be star
companion.  It was discovered in 1998 November, simultaneously by
\rxte\ \citep{chak98} and BATSE \citep{wils98}, and has a pulse period
of 30.95 s.  The source showed two outbursts of $\sim$20 and 40 days
in duration, and the pulsar was spinning up with a very short
timescale of $P/\dot{P} \sim 20$ yr \citep{yoko00}.  The likely Be
companion may be embedded in an \ion{H}{2} region \citep{coe03,coe00}.  Given
a distance of 65 kpc to the SMC, the source had a 0.7--10 keV
luminosity of $\sim$$2 \times 10^{38}$ \ergs.  The \asca\ spectrum is
an inverse, broken power law with iron emission at 6.4 keV, along
with a soft component that was modeled as a broken power law, thermal
bremsstrahlung, or blackbody \citep{yoko00}.  Variations in the soft
X-ray band indicate that the soft component is pulsing.

\rxj\ is another bright transient Be/XBP in the SMC.  It was
discovered as a serendipitous source in a short pointing of \rosat\
\citep{hugh94}.  The \rosat\ plus \asca\ spectrum has luminosity
$\sim$$3\times10^{38}$ \ergs, and the hard component is fitted by a cutoff power
law with $\Gamma \sim 0.4$ \citep{kohn00}. The soft component, for which
$L_{\rm soft}/L_{\rm X} \sim 35\%$, was fitted using either a soft broken
power law in addition to the hard power law or a thin-thermal
emission model \citep[MEKAL;][]{mewe86} with $kT_{\rm MEK} =
0.37$ keV.  Fitting this model required a very low value for overall
metal abundance in the emitting gas ($Z = 0.015$ $Z_{\sun}$), but with a
highly enhanced oxygen abundance in the absorbing column ($Z_{\rm O}
\sim 7$ $Z_{\sun}$).  Unlike the four sources discussed above, its soft component does
not pulsate.

\subsection{Other pulsars with a soft excess}
We know of three other well-studied XBPs that show a soft excess.
\fouru, like Her X-1, is one of the few XBPs with a low-mass
companion.  X-ray pulsations with period 7.7 s were first detected by
{\it SAS 3} \citep{rapp77}, and a faint, blue optical counterpart was
identified by \citet{mccl77}.  There is an unusually short orbital
period of 41.4 minutes \citep{midd81}, which rules out a hydrogen-burning
main sequence star for the companion.  In 1990 June the steady spin-up
of the source suddenly changed to a spin-down state that continues to
the present. The spectrum has repeatedly been fitted with power law
($\Gamma \sim 0.7$) plus blackbody ($kT_{\rm BB} \sim 0.3$ keV) and
emission--line features at $\sim$1 keV
\citep{ange95,owen97,orla98,schu01}.  The pulse fraction is much lower
for soft than hard X-rays \citep{owen97}, suggesting that the soft
component is not pulsing.  \citet{orla98} discovered a cyclotron
absorption feature at $\sim$35 keV, implying that $B \sim 3.5 \times 10^{12}$
G.  They found the broadband (0.1--200 keV) X-ray luminosity to be
$(7.7 \times 10^{34}$ \ergs) $D_{\rm kpc}^2$.  The distance is not well
constrained, so it is difficult to make reliable conclusions about the
physical processes in this source.

There is also evidence for a soft excess in Cen X-3.  It is a
high-mass XPB, with pulse period of 4.8 s and orbital period of 2.1
days \citep{schr72a}, and the O-type companion is estimated to be
at a distance of $\sim$8 kpc \citep{krze74}.  The hard continuum has a
typical cutoff power-law shape, with iron line emission at $\sim$6.5
keV \citep{naga92,burd00} and a cyclotron line feature at
$\sim$30 keV \citep{sant98, burd00, cobu02}, giving $B \sim 3 \times
10^{12}$ G.  Using \bepposax\ data,
\citet{burd00} found a bright pulsating soft excess with a
blackbody shape ($kT_{\rm BB}\sim 0.1$ keV). In this observation Cen X-3
had a large $L_{\rm X}\sim10^{38}$ \ergs.

Vela X-1 also shows a soft excess. It has a long pulse period of 283
s \citep{mccl76} and an orbital period of 8.96 days \citep{form73}.
The power-law spectrum has $\Gamma \sim 1$--1.7 \citep{sako99,choi96}
and the distance is $\sim$1.9 kpc \citep{sada85}, indicating $L_{\rm
X} \sim 10^{36}$ \ergs.  There are cyclotron features at $\sim$24 and
55 keV \citep{krey02,kret96}, indicating
$B\sim 2.4 \times 10^{12}$ G.  The \rosat\ spectrum showed a soft, non-pulsing
thermal bremsstrahlung component with $kT_{\rm tb} \sim 0.5$ keV,
attributed to shock-heated optically thin emission by the strong
stellar wind \citep{habe94b}.  This was similar to the high-mass X-ray
binary 4U 1700--37, which had a remarkably similar \rosat\ soft excess
spectrum \citep{habe94a}.  However, high-resolution grating
spectroscopy of these sources with \chandra\ showed that in eclipse
there were emission lines that likely came from material that was not
collisionally excited but was photoionized by the X-rays from the
neutron star \citep{schu02,boro03}.  We discuss collisionally heated
and photoionized gas emission in \S\S\ 5--6, respectively.

There are two additional transient Be/XBPs in the LMC for which a
spectrally-fitted soft excess has been seen.  \objectname{\exo} was
the brightest source in a survey of the LMC by \citet{habe03}, who
estimated $L_{\rm X}=4.6\times10^{37}$ \ergs.  They fitted the soft
component as thin-thermal (MEKAL) emission with reduced abundances
(0.29 $Z_{\sun}$) of elements heavier than oxygen.  \objectname{\aofive}
was seen in two outbursts with \rosat\ \citep{mavr93}, when the source
had $L_{\rm X} \sim 2$--$4 \times 10^{37}$ \ergs.  They fitted the
spectra with a power law, $\Gamma\sim0.8$, a soft blackbody, $kT_{\rm
bb} \sim 0.2$ keV, and a thermal bremsstrahlung, $kT_{\rm tb} \sim
0.25$, although the spectra have a relatively low number of counts so
there may be considerable uncertainty in these parameters.  The
authors speculate that the blackbody component may be emitted from
near the \alfven\ radius of the neutron star, while the bremsstrahlung
component may be powered by collisional heating of the stellar wind.
Although we mention \exo\ and \aofive\ here, they have little spectral
information in soft X-rays, so we do not discuss them in detail.

\section{Ubiquity of the soft excess}

\begin{deluxetable*}{lccccc}
\tabletypesize{\scriptsize}
\tablecaption{Unabsorbed flux and $N_{\rm H}$ values for a sample of XBPs \label{tbl-1}}
\tablewidth{0pt}
\tablehead{
\colhead{Source}  & \colhead{Flux (\flux)}  &
\colhead{Energy range (keV)\tnm{a}}  &
\colhead{$N_{\rm H}$ ($10^{22}$ \cdens)} &
\colhead{Mission}  &
\colhead{Reference}}
\startdata
& & {\it Soft excess observed} & & \\
Her X-1 & $3.7\times10^{-8\W}$ & 0.3--10\Wp		&  \W0.005 & \xmm & 	1	\\
SMC X-1 & $4.7\times10^{-10}$ & 0.7--10\Wp		&  \W0.5\W\W & \asca & 	2	\\
LMC X-4 & $4.0\times10^{-10}$ & 0.7--10\Wp		&  \W0.057 & \asca & 	3	\\
XTE 0111.2--7317 & $3.6\times10^{-10}$ & 0.7--10\Wp 	&  \W0.18\W & \asca & 4	\\
RX J0059.2--7138 & $6.9\times10^{-11}$ & 0.2--2\W\Wp 	&  \W0.088 & \rosat & 5 	\\
4U 1626--67 & $6.4\times10^{-10}$ & 0.5--10\Wp 		&  \W0.5\W\W & \chandra & 6	\\
Cen X-3 & $5.7\times10^{-9\W}$ & 2.0--10\Wp 			&  \W1.95\W & \bepposax & 7		\\
Vela X-1 & $5.1\times10^{-9\W}$ & 2.0--10\Wp 		&  \W0.86\W & \bepposax & 8		\\
EXO 053109--6609.2 & $1.5\times10^{-10}$ & 0.2--10\Wp 	&  \W0.69\W & \xmm & 9	\\
A 0538--66 & $1.3\times10^{-10}$ & 0.1--2.4 		&  \W0.08\W & \rosat & 	10	\\
& & {\it Soft excess not observed} & & \\				
GRO J1744--28 & $2.3\times10^{-8\W}$ & 2.0--10\Wp 		&  \W5\Wp\W\W\W	& \asca & 	11	\\
GX 1+4 & $8.4\times10^{-10}$ & 2.0--20\Wp 			& 20\Wp\W\W\W	& \asca & 		12	\\
OAO 1657--415 & $4.7\times10^{-10}$ & 2.0--10\Wp 		& 12\Wp\W\W\W			& \bepposax & 	13	\\
4U 1145--619 & $5.0\times10^{-11}$ & 0.2--2\W\Wp 		&  \W0.3\W\W & \exosat & 	14	\\
4U 1907+09 & $4.0\times10^{-10}$ & 2.0--10\Wp 		&  \W1.9\W\W & \asca & 	15	\\
GX 301--2 & $4.6\times10^{-8\W}$ & bolometric 		& 20\Wp\W\W\W	 & \asca & 	16	\\
2S 1417--624 & $1.7\times10^{-10}$ & 2.0--10\Wp 		&  \W1.7\W\W & \einstein & 17	\\
GRO J1008--57 & $2.3\times10^{-11}$ & 0.1--2.4 		&  \W1.3\W\W & \rosat & 18	\\
4U 1145--619 & $5.0\times10^{-11}$ & 0.2--2\W\Wp 		&  \W0.3\W\W & \exosat & 	19	\\
X Persei & $2.0\times10^{-10}$ & 0.1--10\Wp		&  \W0.13\W & \bepposax & 	20
\enddata
\tnt{a}{Energy range over which the flux is defined}
\tablerefs{
   (1) \citealt{rams02};
   (2) \citealt{paul02};
   (3) \citealt{paul02};
   (4) \citealt{yoko00};
   (5) \citealt{hugh94};
   (6) \citealt{schu01};
   (7) \citealt{burd00};
   (8) \citealt{orla98a};
   (9) \citealt{habe03};
   (10) \citealt{mavr93};                 
   (11) \citealt{nish99};
   (12) \citealt{kota99};
   (13) \citealt{orla99};
   (14) \citealt{mere87};
   (15) \citealt{robe01};
   (16) \citealt{endo02};
   (17) \citealt{grin84};
   (18) \citealt{petr94};
   (19) \citealt{mere87};
   (20) \citealt{disa98}.}
\end{deluxetable*}

In this section we address whether the soft excess is a universal
feature of XBPs.  All the brightest sources with low absorption
show this feature, so it is natural to suspect that it may be present
also in other XBPs but not be detected because of either low flux or
high absorbing column.  In Table 2 we show spectrally-determined
$N_{\rm H}$ values and the unabsorbed fluxes for a number of known XBPs.
The sample is taken from Table 1 in \citet{bild97}, plus \exo.
We have included only sources for which there are observations with
sensitivity at the energies of the soft excess less than 2 keV.  The fluxes
are defined in different bands in the range 0.1--20 keV, so there
is some error (factors of a few or so) in the relative fluxes from
these sources, although this does not affect the overall pattern observed.

The sources in the sample are plotted as a function of flux and
$N_{\rm H}$ in Fig. 1.  The XBPs with a known soft excess are shown as
open stars; those without as solid squares.  It is clear that the brighter
systems with less absorption are far more likely to show a soft
excess, and that no sources with high absorption ($N_{\rm H} > 3
\times 10^{22}$ \cdens) show this feature.  There is an apparent
boundary between the sources that show a soft excess and those that do
not, suggesting that the observability of this feature depends on
flux and $N_{\rm H}$.

Marked with an asterisk is the Be/XBP X Persei, which has a markedly
different soft excess from the other sources.  \citet{cobu01} modeled
this component as a blackbody with $kT_{\rm BB} \sim 1.4$ keV, which
is an order of magnitude higher than the typical temperature.  X Per
is very nearby ($D \sim 0.7$ kpc) and under-luminous ($L_{\rm X} \sim
10^{34}$ \ergs) \citep{disa98}, and the authors attributed the soft
component to blackbody emission from the accreting polar cap.

\begin{figure}
\centerline{\includegraphics[width=3.5in]{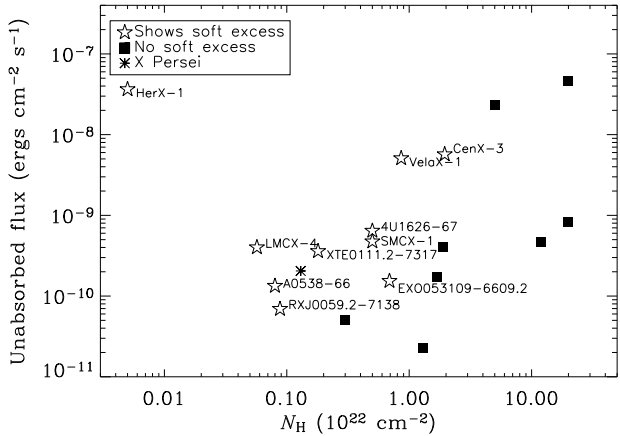}}
\caption{Observed $N_{\rm H}$ and unabsorbed
fluxes for the XBPs in Table 1 of \citet{bild97}, plus \exo.  Sources with a
known soft excess are shown as stars, while those without are shown
as squares.  The low-luminosity source X Per is shown as an asterisk
(see text for details).}
\end{figure}

\begin{figure}
\centerline{\includegraphics[width=3.5in]{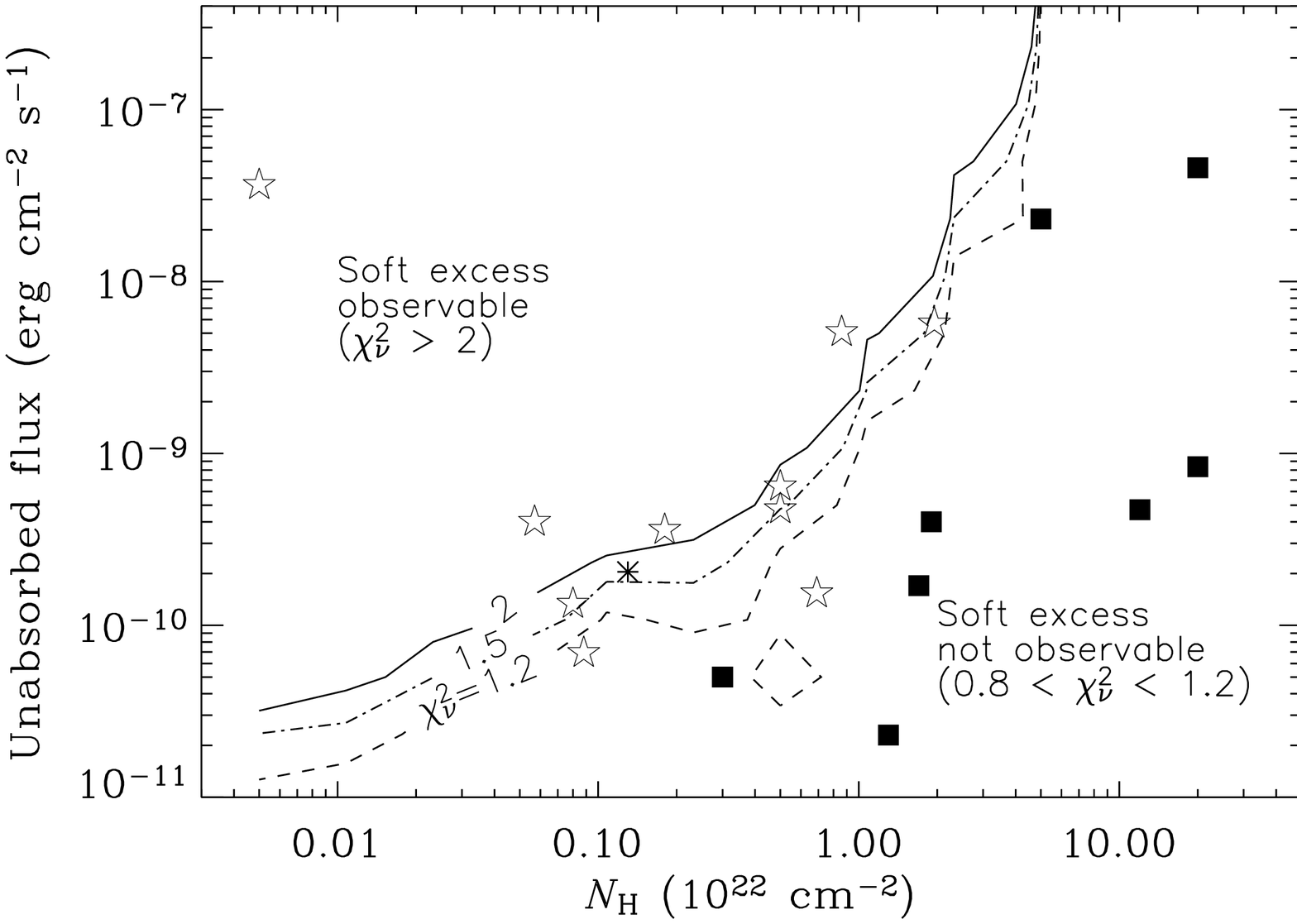}}
\caption{Observability of the soft excess as a function of $N_{\rm H}$
and unabsorbed fluxes, from fits to simulated \rosat\ data.  Contours
of $\chi_{\rm \nu}^2$ for fits to a simple absorbed power law are shown.
The soft excess is observable for high $\chi_{\rm \nu}^2$ and not
observable for low $\chi_{\rm \nu}^2$.  The XBPs from Fig. 1 are also
shown for comparison; note that the $\chi_{\rm \nu}^2$ contours are
similar in shape to the boundary between observed XBPs with and
without a soft excess.}
\end{figure}

We modeled the detectability of the soft excess at various fluxes and
$N_{\rm H}$ by performing fits to simulated data.  The response of the
\rosat\ PSPC detector \citep{pfef86} is used, because many XBPs have
been observed with this detector and because it is sensitive in the
soft X-ray band.  Using XSPEC \citep{arna96}, we created simulated
PSPC spectra of a source with a spectrum like that of SMC X-1 ($\Gamma = 0.9$,
$kT_{\rm BB}=0.15$ keV, and $L_{\rm soft}/L_{\rm X}=0.04$), and varied
$N_{\rm H}$ and the overall unabsorbed flux in the 0.2--3 keV band.
The simulated exposure time was 25 ks, and counting statistics were
included.  We also included a typical background spectrum, taken from
the PSPC observation of SMC X-1 from 1991 October.  This background
has a surface brightness in the PSPC band (0.1--2.4 keV) of
$2.1\times10^{-2}$ counts \pers\ arcmin$^{-2}$, which is similar to
those found for Cen X-3, LMC X-4, and Her X-1.  More details on PSPC
observations of these sources are given in \S\ 4.1.

\begin{figure*}
\plotone{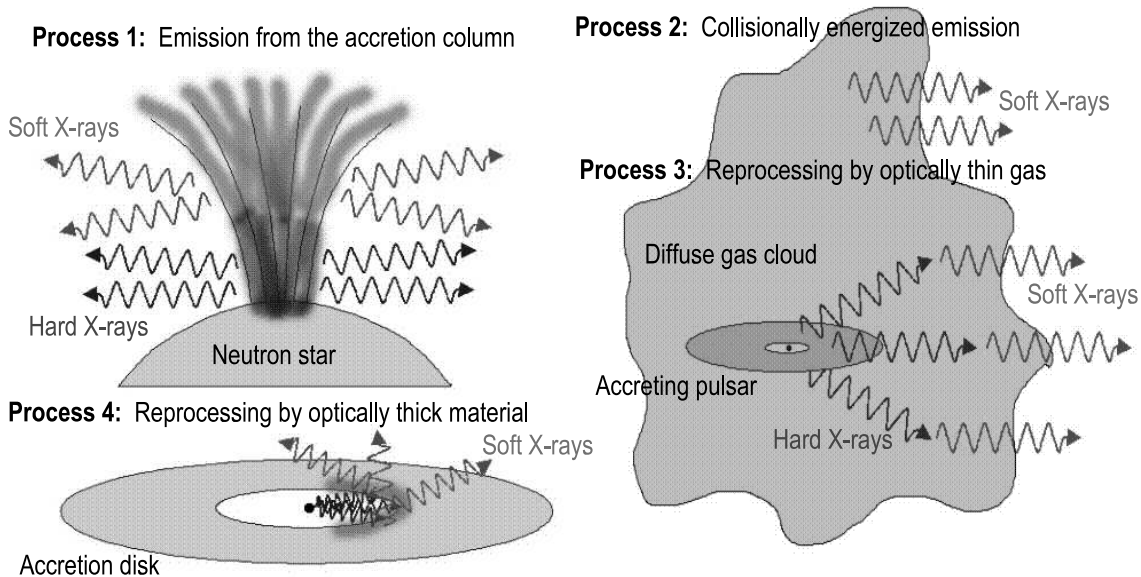}
\caption{Schematic showing the four soft excess emission processes
considered in this paper.}
\end{figure*}

Next, we determined whether each spectrum, after background subtraction,
could be fitted with an absorbed power law without a soft excess component.
For the fits we took initial values of $N_{\rm H}$ and $\Gamma$ and
overall normalization from the simulated data, and allowed all these
parameters to vary.  We recorded $\chi^2_{\rm \nu}$ for the best fit to
each simulation.  For low $\chi^2_{\rm \nu}$, we conclude that a power law
spectrum satisfactorily fits the simulated data and so the soft excess
is {\it not detectable}.  For high $\chi^2_{\rm \nu}$ the soft excess is
detectable.

The fit results are plotted in Fig. 2, along with the XBPs from
Fig. 1.  Contours of $\chi^2_{\rm \nu}=1.2$, 1.5, and 2 are shown.  For
low flux and high $N_{\rm H}$ (lower right), $\chi^2_{\rm \nu}$ varies
somewhat randomly between 0.8 and 1.2; for clarity, these contours are
not shown. We find that the soft excess is detectable in sources with
high flux and low $N_{\rm H}$, and undetectable for low flux and high
$N_{\rm H}$.  The steep contours of $\chi^2_{\rm \nu}$, which indicate the
boundary of detectability, are close to and similar in shape to the
actual boundary seen in observations.  This suggests that many sources
with low flux or high $N_{\rm H}$ do in fact have a soft component
but that this feature has not been observed because of selection
effects.  We conclude that the soft excess is probably a very common
intrinsic feature of XBPs.

As discussed in \S\ 2, observations of the soft excess in XBPs have
been modeled in a variety of ways, many of which require different
emission processes. Given the similarities of many of the observed
soft components and the apparent ubiquity of this feature, it is
likely that it has a common origin in many sources. Possible
mechanisms can be separated into four main types:

\begin{enumerate}
\im emission from the accretion column,
\im thermal emission from collisionally energized, diffuse gas around the neutron star, 
\im reprocessing of the hard (power law) X-rays by diffuse material
around the neutron star, and  
\im reprocessing by optically thick, dense material around the neutron star.
\end{enumerate}

In the following sections, we consider these four scenarios and
examine them in light of observational data and the physics of the
relevant emission.  Simple schematics of the various models are shown
in Fig. 3.  For a model to be viable it must satisfy two constraints:

\begin{enumerate}
\im  the energy balance of the system must be able to
produce the observed soft luminosity and
\im the spectrum, pulsation properties, and other observed features must be
consistent with model predictions.  
\end{enumerate}

\section{Emission from the accretion column}
It is natural to suspect that the soft component originates in the
same region as the hard component, that is, in the column of accreting gas.
Low-luminosity soft emission from the polar cap may exist in X Per
\citep{disa98}, and the more luminous soft components in SMC X-1, LMC
X-4, and \xtej\ have been modeled by non-thermal broken power-law
emission from the accretion column \citep{yoko00,paul02}.  This model
predicts soft pulses, because emission is confined to a small region
that rotates with the neutron star.

\subsection{Brightness temperatures of accretion columns}
In XBPs the accreting material flows along magnetic field lines from
the magnetosphere, at radius $R_{m}$, to the surface of the
neutron star at $R_{\rm NS}$.  To estimate the surface area of the
accretion column, we follow \S\ 6.3 of \citet{fran02}.  We assume that the
magnetic field is a dipole that is tilted by an angle $\alpha$ with
respect to the accretion disk, and that only field lines that pass
through the disk outside $R_{m}$ accrete.  It is
straightforward to show that all the accreting field lines intersect the
neutron star inside an angle $\beta$ from the pole, where $$\sin^2{\beta}=(R_{\rm NS}/R_{m})\sin^2{\alpha}.$$
Thus the radius of the accreting polar cap is 
$$R_{\rm col} \simeq R_{\rm NS} \sin{\beta} \sim R_{\rm NS}(R_{\rm NS}/R_{m})^{1/2},$$
assuming $\sin{\alpha} \sim 1$.
For the XBPs we consider, $R_{m} \sim 10^8$ cm so for $R_{\rm NS} \sim 10^6$ cm, 
$R_{\rm col} \sim 0.1 R_{\rm NS}$. If we picture the accretion column as a
cylinder with height $\lesssim R_{\rm NS}$, the surface area of the
column is
$$A_{\rm col} \lesssim 2 \pi R_{\rm col} R_{\rm NS}= f_{\rm col} 4 \pi R_{\rm NS}^2.$$
Here $f_{\rm col}$ is simply the surface area of the accretion column as a
fraction of the neutron star's surface area, with $f_{\rm col}\lesssim 0.1$.

For high $L_{\rm soft}$, this small area implies a very high
temperature for the emitting gas.  The minimum temperature for
material radiating a spectral luminosity $L_{\rm \nu}$ in a given band is
the brightness temperature $T_b$, which is the temperature of a
blackbody that would produce the same $L_{\rm \nu}$.  For an emitting area
$A$, $$L_{\rm \nu}=A\pi B_{\rm \nu}(T_b).$$ The value of $T_b$ for
different energy ranges sets a lower limit on the temperature of the
gas emitting at those energies.

\begin{deluxetable}{llrcc}
\tabletypesize{\scriptsize}
\tablecaption{\rosat\ PSPC data used for $T_b$ calculations}
\tablewidth{0pt}
\tablehead{
 \colhead{Source} & \multicolumn{2}{c}{Date/Duration (ks)} &  \colhead{$N_{\rm
H}$ ($10^{22}$ \cdens)}}
\startdata
Her X-1	& 1993 Aug& 5 & 0.003 \\
SMC X-1	& 1991 Oct& 17 & 0.36\W \\
LMC X-4	& 1991 Oct--Nov& 33 & 0.05\W \\
\rxj\	& 1993 May& 5 & 0.04\W \\
Cen X-3	& 1992 Jan& 9 & 2\Wp\W\W\W \\
Vela X-1& 1992 Nov& 17 & 0.5\W\W 
\enddata
\end{deluxetable}

We used data from the \rosat\ PSPC observations for Her X-1, SMC X-1
LMC X-4, and Cen X-3, as well as \rxj\ and Vela X-1, which we include
for comparison even though their lack of soft pulsations would appear
to rule out accretion column emission.  Data on these observations are
given in Table 3.  The PSPC is a proportional counter detector with
response at low energies, from 0.1 to 2.4 keV, and with spectral
resolution $\Delta E/E=0.43(E/0.93 \mathrm{\ keV})^{-1/2}$
\citep{pfef86}.  We obtained the data from the NASA HEASARC archive
and used the standard event-screening criteria.  We extracted the
source and background spectra from circular and annular regions
centered on the source. In the case of \rxj, a nearby supersoft source
(1E 0056.8--7154) was excluded from the selection regions, as
described in \citet{kohn00}.

We assume that {\it all} the observed flux in the spectrum comes from
a column of area $A_{\rm col}$ and determine $T_b$ as a
function of energy.  We use energy bins corresponding to the PSPC
resolution and fit each bin with a power law of constant energy
(photon $\Gamma = 1$), modified by typical values of $N_{\rm H}$
(shown in Table 3). Each XSPEC fit gives the normalization of the
power-law $I_{\rm pl}$ (in photons cm$^{-2}$ s$^{-1}$ keV$^{-1}$ at 1
keV) and its error.  The energy flux in each bin (from $E_1$ to $E_2$)
is then
$$F_{\rm obs}=\int_{E_1}^{E_2}E (I_{\rm pl} E^{-1})dE=I_{\rm pl}
(E_2-E_1).$$ Given distances listed in Table 1, we calculate the
luminosities and then $T_b$ in each bin.  To obtain a lower
limit on $T_b$, we use the $3 \sigma$ lower limit on $I_{\rm
PL}$.

The isotropic luminosity observed in a given bin is
$$L_{\rm obs}=4 \pi D^2 F_{\rm obs},$$   and the luminosity {\it emitted} from
the accretion column in that bin is
$$L_{\rm em}=A_{\rm col} \pi \int_{E_1}^{E_2}B(T_b)dE.$$

\begin{figure}
\centerline{\includegraphics[width=3.5in]{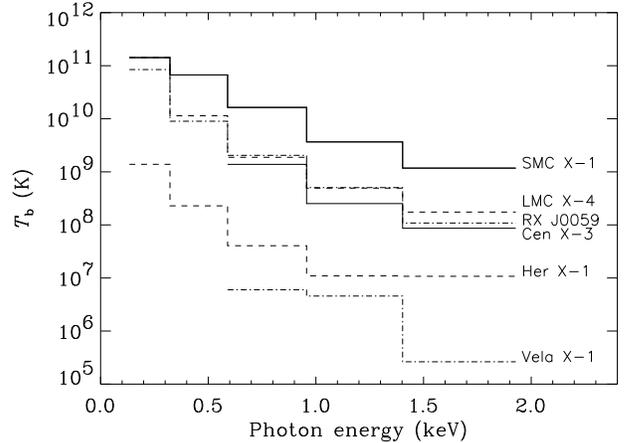}}
\caption{Brightness temperatures for the observed fluxes ($3\sigma$
lower limit) from \rosat\ PSPC data for a sample of XBPs.  We
assume an emission region with area a fraction $f_{\rm col}=0.1$ of
the surface area of the neutron star ($R_{\rm NS}=10$ km). For $E \ll
T_b$, $T_b \propto f_{\rm col}^{-1}$.}
\end{figure}

Letting $L_{\rm obs}=L_{\rm em}$, and using $A_{\rm col}=f_{\rm col} 4 \pi R_{\rm NS}^2$, we have
$$\int_{E_1}^{E_2}B(T_b)dE=\frac{D^2}{\pi R_{\rm NS}^2 f_{\rm
col}} F_{\rm obs}.$$ We solve this numerically to find $T_b$.
In the Rayleigh-Jeans limit $E \ll kT_b$ this gives, for $f_{\rm col}=0.1$,
$R_{\rm NS}=10$ km, $E_1$ and $E_2$ in keV, and $L_{38}$, the
luminosity in the bin in $10^{38}$ \ergs,
$$T_b=1.75\times10^{10} \ \textrm{K} \ \frac{L_{38}}{E_2^3-E_1^3}.$$
For each source we plot the
values of $T_b$, given $f_{\rm col}=0.1$, in Fig. 4.  Vela X-1
and Cen X-3 are highly absorbed below 0.5 keV, so we do not calculate
$T_b$ for those energies.

As expected, $T_b$ is higher for more luminous sources.  For low
energies, we find that for SMC X-1, LMC X-4, and \rxj, the observed
flux gives $T_b \gtrsim 10^{11}$ K, even for a relatively large
accretion column area ($f_{\rm col}=0.1$).  At higher energies, we have
$T_b>10^8$ K (and, remarkably, $T_b>10^9$ K for SMC X-1).
These $T_b$ values represent the minimum possible temperatures
of the gas emitting at these energies.  Are these temperatures
consistent with models of accretion columns?  In general, such models
have been separated into two types, depending on the shock mechanism
that decelerates the gas as it approaches the neutron star's surface.
The shock can be dominated by gas pressure (for low $L_{\rm X}$
systems) or by radiation pressure (for $L_{\rm X} > 10^{36}$ \ergs).
The latter models are applicable for the luminous XBPs.  This type of
accretion column consists of a mound of subsonic, settling material
below a radiation-dominated shock front \citep{beck98}.
\citet{burn91} modeled the dynamics and emission properties of the
accretion mounds, and found that temperatures in the mound vary from
$10^8$ K at the outer boundary to $2\times10^8$ K at large optical depths.

\citet{burn91} note that the accretion mound emission is
Compton-scattered by the overlying shock, which could increase $T_{\rm
b}$.  In principle, if the gas free-falls all the way to the neutron
star's surface and is stopped in a thin shock, then temperatures of
$\sim$$10^{12}$ K are possible \citep{shap83}.  However, we expect
that collisions in the column would slow the gas, and it is expected
that a radiative shock is not perfectly thin but has width
approximately equal to the
radius of the polar cap \citep{beck98}.  Therefore we expect the shock
temperature to be lower.  It is difficult to see how Comptonization by
a real shock could produce the very high soft fluxes observed.

This result raises the interesting, puzzling question of how, for SMC X-1, even
the hard emission above 1.5 keV could be produced by the accretion
column.  For a large column size $f_{\rm col}=0.1$, the $T_b
\sim 10^9$ K is hotter than models predict.  This suggests that either
the column is larger in extent (i.e. $f_{\rm col}> 0.1$) or it is more
energetic ($T \gtrsim 10^9$ K) than predicted.  This
question is worth investigating in future work.

\begin{figure}
\centerline{\includegraphics[width=3.5in]{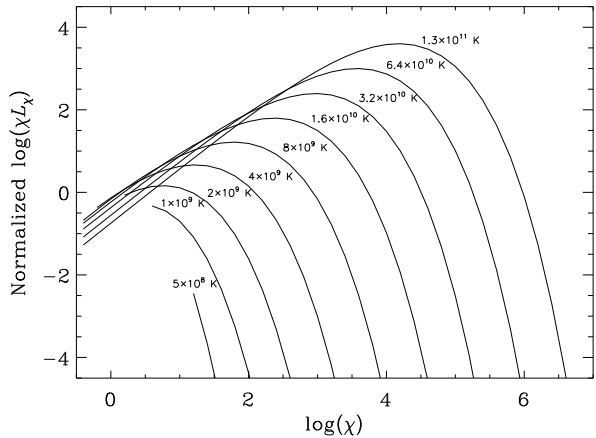}}
\caption{Spectra of cyclo-synchrotron radiation for different
temperatures, taken from \citet{maha96}.  Spectra are plotted as a
function of $\chi=E/E_{\rm B}$.}
\end{figure}

\subsection{Comparison with cyclo-synchrotron spectra}
Even if accretion column temperatures are very high, it is difficult
to produce the observed soft spectral shapes.  For high $T_b \gtrsim
10^7$ K, blackbody or thermal bremsstrahlung emission would produce
the majority of the flux at energies $\sim kT > 1$ keV, which is
inconsistent with observations.  One possible way out is to note that,
in the large magnetic field of the accretion column, the dominant
emission process is likely to be cyclo-synchrotron radiation.  If this
radiation is thermal, the peak of the emission can be at much lower
energies than $kT$. The shape of this spectrum has been calculated in
rigorous detail by \citet{maha96}, who show that it is a function of
$\chi=E/E_{\rm B}$, where
$$E_{\rm B}=\frac{heB}{2\pi m_e c}=11.6 (B/10^{12} \rm{\ G}) \rm{\ keV}$$
is the (non-redshifted) cyclotron energy.  Therefore, the energy range
of cyclo-synchrotron emission depends largely on $B$.  We plot the
calculated spectra from \citet{maha96} in Fig. 5.

At $T=3.2\times10^{10}$ K, which is smaller than the minimum $T_{\rm
b}$ that we obtained for the luminous XBPs at low energies, the spectrum
peaks at $\log{\chi} \sim 3$.  A typical soft component in the
luminous XBPs has the peak of $\nu F_{\rm \nu}$ at $\sim 4 kT_{\rm BB}
\sim 0.7$ keV.  For cyclo-synchrotron emission to peak at this energy,
we require that $$E_{\rm B} \sim 0.7 \rm{\ keV}/\chi \sim 7 \times 10^{-4} \rm{\
keV},$$ implying that $B \sim 6\times10^7$ G.  For a dipole field with
surface strength $4 \times 10^{12}$ G, $$B(r) \sim
4\times10^{12}(r/R_{\rm NS})^{-3},$$ implying that the emission must
come from a radius $r \sim 40 R_{\rm NS}$.  This is much larger than
the expected size of the accretion column, and the gas cannot attain
such a high temperature so far from the neutron star.

It is difficult to conceive of any other emission mechanism that could
produce a high soft luminosity from a small area and yet have the
spectrum cut off at the energies observed.  We conclude that in the
luminous XBPs (SMC X-1, LMC X-4, \rxj, \xtej, and Cen X-3), the soft
component must be emitted from a region {\it larger than the surface
of the neutron star}.

\section{Emission by a collisionally energized cloud}
A large, diffuse cloud of gas around the neutron star would have the
size needed to produce the high $L_{\rm soft}$ and might be powered
by collisional energy. Such emission may exist in Vela X-1 and 4U
1700--37 \citep{schu02,boro03}, and perhaps also in RX J0101.3--7211
and AX J0103--722, which were observed in the \xmm\ survey of the SMC by
\citet{sasa03}.  These sources have $L_{\rm X} \sim 10^{35}$ \ergs,
and were fitted with a power law plus an optically thin thermal emission
model (MEKAL) with $kT_{\rm MEK} \sim 0.2$ keV.

Collisional heating is expected in XBP systems; a detailed
hydrodynamic simulation of LMC X-4 by \citet{boro01} found shocks and
wake structures that could heat the gas and cause optically thin
X-ray emission \citep{fran80}.  Fitting a thin thermal emission model
to data from SMC X-1 and LMC X-4, \citet{paul02} found that the
observed high $L_{\rm soft}$ requires $n^2V \sim 10^{61}$, which for
typical gas density $n < 10^{12}$ \vdens\ implies a very large
emission region of $R_{\rm cloud} > 10^{12}$ cm.  The same is true for
\xtej\ and \rxj, which have similar soft spectra.

The large emission region immediately rules out this process for
sources with a soft component with short pulsations, such as SMC X-1
and \xtej.  A pulse period of about a few seconds is much shorter
than the light-crossing time of such a large cloud, and so the cloud's
emission could not pulsate.  For SMC X-1 and LMC X-4, the size of the
emitting region is also constrained by the drop of the soft flux
during eclipses \citep{vrti01,boro01}.  The eclipse light curves for
SMC X-1 and LMC X-4 \citep[e.g.,][]{woo95, laba01} show that the soft
flux falls with a timescale $\lesssim 10\%$ of the total eclipse
duration.  Thus, $R_{\rm cloud} < 0.1 R_{\star} \lesssim 10^{11}$
cm. Such arguments rule out collisionally energized emission for SMC
X-1, LMC X-4, and \xtej, but the process may still be important in
\rxj, which shows neither eclipses nor pulsation at soft energies.

In XBPs the main input of kinetic energy is from a wind, simply
$\dot{E}=\dot{M}v_{\rm w}^2/2$, where $v_{\rm w}$ is the wind velocity,
typically $\sim$2000 km \pers.  Mass loss rates from the star and from
the irradiated accretion disk are typically $\dot{M} \lesssim 10^{-6}\
\msun$ yr$^{-1}$, giving $\dot{E} \lesssim 5 \times 10^{36}$ \ergs.
Such emission can easily power the soft components of lower luminosity
XBPs such as Vela X-1.  However, it is possible for the more luminous
soft excess of \rxj\ only if there is a very vigorous wind for which
{\it all} the kinetic energy is converted to soft X-rays.  This is
unlikely.

In addition, diffuse thermal emission in \rxj\ should show line
emission from metals. In their MEKAL fit to the \asca\
spectrum, \citet{kohn00} obtain an extremely low $Z=0.02$, which fits
best because it greatly reduces the flux of most X-ray lines.  While
the metallicity of the SMC is expected to be low, this value is
unrealistic.  It is more likely that the non-detection of lines is due
to optically thick (blackbody-like) emission, which we discuss in \S\
7.

We conclude that collisionally energized diffuse emission can
power the soft excess only in systems with $L_{\rm X} \lesssim 10^{36}$
\ergs.  This may explain part of the soft components in Vela X-1 and
4U 1700--37, but in \rxj\ the soft excess is too luminous for this
process.  With high spectral resolution observations, it is possible
to easily identify this type of emission by the presence of X-ray
emission lines. In addition there should be other spectral signatures
from diffuse gas around XBPs, because of {\it photoionization} of the gas
by the central X-ray source.  We discuss this process in the next section.

\section{Reprocessing by a diffuse cloud}
It is possible that the soft flux arises from {\it reprocessing} of
the hard X-rays by a diffuse cloud of gas around the neutron star.  Such
a cloud must absorb and re-radiate a fraction of the incident hard
luminosity equal to $f_{\rm rep} = L_{\rm soft}/L_{\rm X} \sim 0.1$
and must produce an output spectrum like those observed in
XBPs. Because of the high $L_{\rm X}$ in these systems, the gas is highly photoionized by the X-ray flux.  Emission from such a gas
may contribute to the soft excess in Vela X-1 and 4U 1700--37
\citep{boro03, schu02}. The XSTAR code is designed to calculate the
detailed structure of photoionized gas, and has been used extensively
to model the circumstellar environments in XBPs.  For details on the
code and its applications, see \citet{kall01}.

XSTAR works by determining the temperature, ionization and excitation
structure of a spherical gas cloud illuminated by a central point
source of X-rays.  It calculates the absorption by the gas due to
photoionization and Compton heating and the emission from the gas through
recombination, free-free emission, and atomic line emission. In the
case of an optically thin low-density ($< 10^{12}$ \vdens) gas, the
structure is a function simply of the ionization parameter
$\xi=L/nR^2$, where $L$ is the luminosity of the central source, $n$
is the number density of the gas, and $R$ is the distance from the source
\citep{tart69}.

XSTAR has been used by \citet{wojd00} and \citet{boro01} to compare
the observed eclipse spectra from SMC X-1 and LMC X-4 with hydrodynamic
simulations.  \citet{vrti01} used XSTAR to analyze relatively
low-luminosity ($\sim$$10^{35}$ \ergs) photoionized line emission from
SMC X-1 seen in \chandra\ ACIS spectra.  In eclipse and in the low
state (when the neutron star was blocked by the disk), they found
lines that correspond to XSTAR simulations with $1 < \log{\xi} <
1.5$.  \citet{vrti01} found that at a radius of $2\times10^{12}$ cm
from the neutron star, these values for $\xi$ indicate a gas density
of $\sim$$10^{12}$ \vdens.

For most of the XBPs with $L_{\rm X} \gtrsim 10^{38}$ \ergs, we do not
expect a diffuse cloud to be the source of the soft excess as
discussed in \S\ 5.  However, such a scenario is possible in \rxj,
which does not show soft pulsations or eclipses, and so we test
the possiblity that optically thin reprocessing is the main source of
the soft excess in this system.  We have used the latest version of
XSTAR, version 2.1k, to simulate the photoionized gas in this source.

\subsection{XSTAR input}

\begin{figure}
\centerline{\includegraphics[width=3.5in]{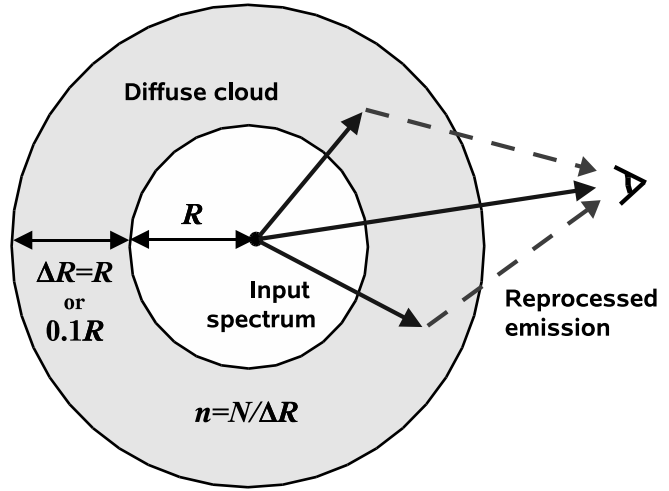}}
\caption{Schematic of the XSTAR models.  The observed emission is
taken to be the sum of the input power law and the outward emitted
spectrum.}
\end{figure}

Because our models must reprocess a substantial fraction of the
incident X-ray flux, we required that the intervening gas subtend a
large solid angle around the central source.  We thus modeled the
reprocessing by spherical shells of gas centered on the neutron
star. For a cloud of radial extent less than $10^{11}$ cm, column densities
higher than $10^{23}$ \cdens\ imply densities higher than $10^{12}$ \vdens, for which we
cannot parameterize our models simply in terms of $\xi$.

Instead, we constructed a series of models of gas in spherical shells
with varying radius $R$ and column density $N$ (Fig. 6).  Because the
gas density in the system falls off with radius, the largest clouds of
constant density should have thickness comparable to the scale length,
or $\Delta R \sim R$.  We therefore ran models with $\Delta R=R$ and $0.1R$ to explore the effects of higher density parcels.

\begin{deluxetable*}{llcccccccc}
\tablecolumns{9}
\tabletypesize{\scriptsize}
\tablecaption{Results of selected XSTAR runs\tnm{a}}
\tablewidth{0pt}
\tablehead{
\colhead{} & \colhead{} & \colhead{} & \colhead{} & 
\multicolumn{2}{c}{$Z=0.25$} & \multicolumn{2}{c}{$Z=0.01$} & 
\multicolumn{2}{c}{$Z=1$} \\
\colhead{$\log{R}$}  & \colhead{$\log{N}$} & \colhead{$\log{n}$} &
\colhead{$\log{\xi}$} & \colhead{$\log{T}$} & \colhead{$f_{\rm rep}$} &
\colhead{$\log{T}$} & \colhead{$f_{\rm rep}$} &
\colhead{$\log{T}$} & \colhead{$f_{\rm rep}$}}
\startdata
& & & & &  $\Delta R=R$ & & \\	
\W9.0 & 24.0 & 15.0 & 4.7--5.4 & 7.6--7.6 & $1.0\times10^{-4}$ & 7.6--7.6 &
$1.3\times10^{-5}$ & 7.6--7.6 & $3.8\times10^{-4}$ \\
10.0 & 24.0 & 14.0 & 3.7--4.4 & 6.9--7.5 & $3.6\times10^{-3}$ & 7.0--7.5 &
$4.9\times10^{-4}$ & 6.9--7.5 & 0.014 \\
11.0 & 23.5 & 12.5 &3.2--3.9 & 4.6--7.2 & 0.045     & 4.6--7.2 & 0.099
& 6.4--7.2 & 0.018 \\
11.0 & 24.0 & 13.0 &2.7--3.4 & 4.7--6.5 & 0.14\W      & 4.5--5.6 & 0.022
& 6.2--6.6 & 0.14\W \\
11.5 & 23.5 & 12.0 & 2.7--3.4 & 4.6--6.4 & 0.087     & 4.3--5.1 & 0.040
& 5.2--6.6 & 0.099 \\
12.0 & 23.0 & 11.0 & 2.7--3.4 & 5.4--6.5 & $8.7\times10^{-3}$ & 4.8\tnm{b} &
$7.7\times10^{-3}$     & 6.2--6.6 & 0.022 \\
& & & & &  $\Delta R=0.1R$ & & \\	
11.0 & 24.0 & 14.0 & 2.3--2.4 & 4.5--5.1 & 0.21\W & 4.5--4.9 & 0.045 & 4.4--5.8 & 0.41\W\\
11.5 & 23.5 & 13.0 & 2.3--2.4 & 4.8\tnm{b} & 0.14\W & 4.7\tnm{b} & 0.044 & 4.6\tnm{b} & 0.26\W\\
12.0 & 23.0 & 13.0 & 2.3--2.4 & 4.8\tnm{b} & 0.061 & 4.5\tnm{b} & 0.051 & 4.6\tnm{b} & 0.14\W
\enddata
\tnt{a}{All quantities are in cgs units.}
\tnt{b}{Model run with constant temperature throughout cloud.}
\end{deluxetable*}

For the input spectrum we used an exponentially cutoff power law
that matches the hard component of \rxj\ as fitted by
\citet{kohn00}.  The spectrum has luminosity $L_{\rm X} = 2.6\times
10^{38}$ \ergs, $\Gamma=0.43$, $E_{\rm cut}=6.4$ keV, and $E_{\rm
fold}=9.3$ keV.  For gas composition we used a standard SMC value of
$Z=0.25 $ \citep{welt01}, as well as $Z=0.01$ and 1.

There are large turbulent velocities in the gas around the neutron
star, so we modified the standard XSTAR code to include these
effects in the line opacity.  For each model we set the turbulent
velocities equal to the Keplerian velocity, $$v_{\rm turb}=v_{\rm
K}=1.9\times 10^4 (R/10^8\rm{\ cm})^{-1/2} \rm{\ km\ s}^{-1}$$ for a
neutron star mass of $1.4 \ M_{\sun}$.  This decreases the line
opacities by a factor of $v_{\rm thermal}/(v_{\rm turb}\sqrt{A})$ where
$A$ is the atomic mass.

\begin{figure*}
\plotone{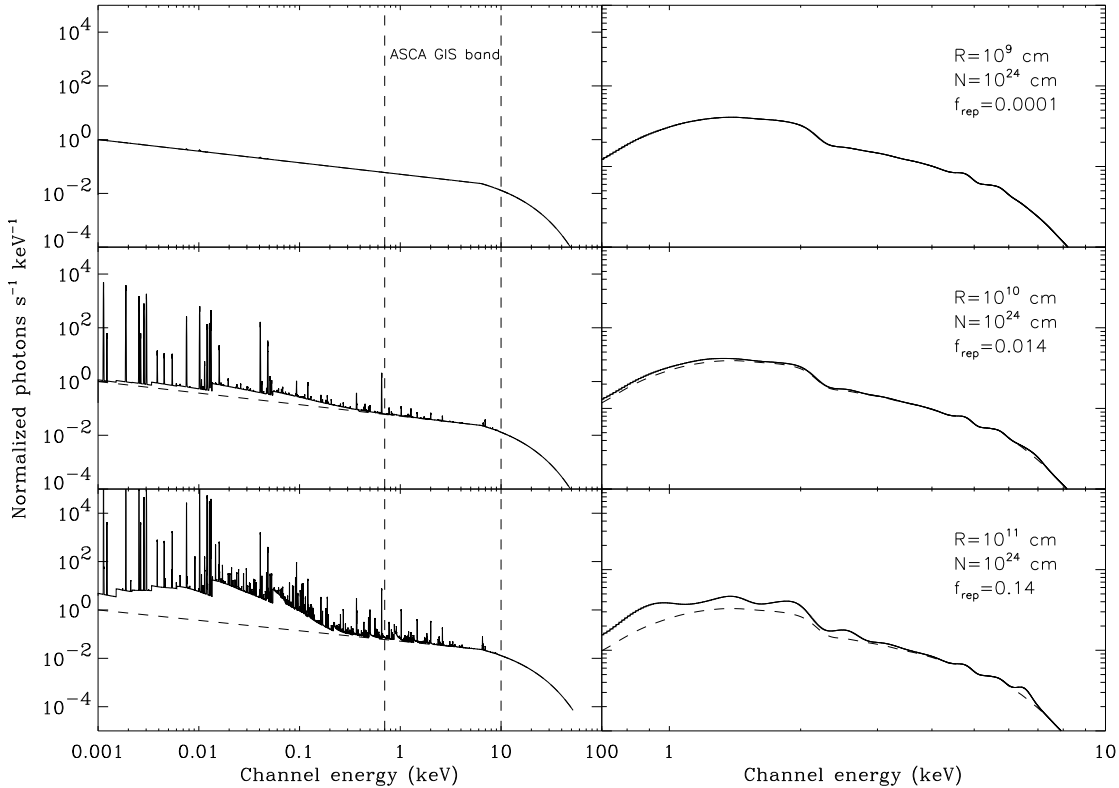}
\caption{\textit{Left:} Sample photon spectra from XSTAR showing the
effects of increased reprocessing with larger radius.  The models have
$Z=0.25$ $Z_{\sun}$, $N=10^{24}$ \cdens, and $\Delta R=R$.  The dotted
line shows the input spectrum, matching the exponentially cutoff power
law for \rxj.  Note the recombination continuum at energies below 0.1
keV.  \textit{Right:} Part of each spectrum for 0.7--10 keV, convolved
with the response of the \asca\ GIS2 detector.  Note that features
become visible in the \asca\ spectrum as reprocessing increases.}
\end{figure*}

We ran models with radii from $10^8$ cm (the expected radius of
the inner disk) to $10^{12}$ cm (a few times the size of the binary
system during outburst which is $R\sim R_{\star} \sim 5 R_{\sun}$).  Most of
the reprocessing column should be contained within this radius.  Close
to the neutron star, the accretion flow can create large gas
densities and columns $\sim$$10^{24}$ \cdens\ or greater.  At $R
\gtrsim R_{\star}$, the wind from the Be star is expected to
blow away the surrounding interstellar medium so that the gas density is
determined mainly by the star's mass loss.  For a spherical wind,
$$\dot{M}=n \mu m_{\rm H} 4 \pi R^2 v_{\rm w}$$ where $\mu$ is the
molecular weight, $\sim$1.4. For $\dot{M}\lesssim10^{-6}$ M$_{\sun}$
yr$^{-1}$ and $v_{\rm w}\gtrsim1000$ km \pers, we have upper limits in
our models of $N \lesssim 10^{23.5}$ and $10^{23}$ \cdens, for
$R=10^{11.5}$ and $10^{12}$ cm, respectively.  We note that while large
densities may be possible at smaller radii, we did not run models with
$N > 10^{24}$ \cdens\ because the effects of Compton scattering are
not explicitly included in XSTAR.  We discuss Compton-thick
reprocessing in the next section.

In most models, we allowed XSTAR to vary the temperature in each
subshell to minimize heating minus cooling as it stepped outward through
the cloud.  However, for some runs at large radii
($10^{11.5}-10^{12}$ cm), the model encountered problems converging on
the temperature.  In these cases we set the entire cloud to be at the
single temperature that minimized heating minus cooling over the entire
cloud.
\vspace{0.4in}

\subsection{XSTAR results}

\begin{figure}
\centerline{\includegraphics[width=3.2in]{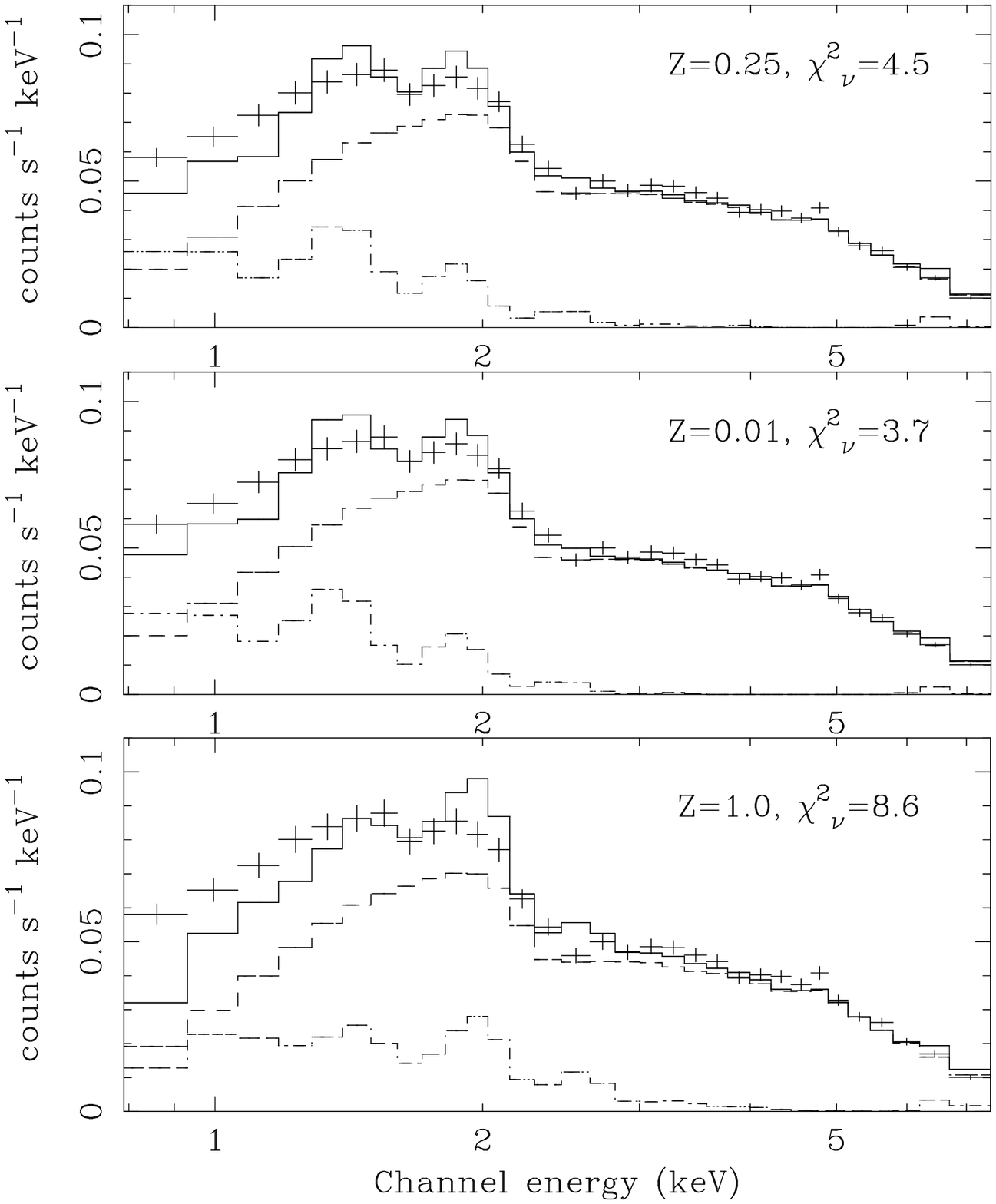}}
\caption{Observed \asca\ GIS2 spectrum for \rxj, fitted with the input
power law plus the six most luminous outward emission spectra from
XSTAR, for $Z=0.25$, 0.01, and 1 $Z_{\sun}$. Each fit had 24 degrees of freedom,
and the best-fit $\chi^2_{\rm \nu}$ values are as shown.  The dashed
lines indicate the contributions from the power law and the
reprocessed emission.  Note that even when emission from multiple
models is included, the spectra do not reproduce the observations.}
\end{figure}

Outputs of sample model runs are given in Table 4.  Because a simple
power law fits the hard end of the spectrum well, there is likely to
be little absorbing gas in the system directly along the line of
sight, so the observed spectrum is the sum of the input hard power
law plus the reprocessed emission from surrounding gas.  Samples of
these spectra are shown in Fig. 7.

We find that none of the models are able to reproduce the observed
emission from \rxj.  In models in which $\log{\xi} \gtrsim
5$ (that is, for small radii) there is virtually no reprocessing for
any column density, because the gas temperature is very high ($\sim$$5
\times 10^7$ K), all the atoms in the gas are completely ionized, and
therefore, the opacity to photoionziation interactions is
low. Reprocessing inside this radius requires densities {\it much
higher} than those that are possible to model using XSTAR.  This approaches
dense, optically thick reprocessing, which is discussed in the next
section.

For larger $R \ge 10^{11}$ cm (and thus smaller $\xi$), the spectrum is
markedly altered by the photoionized gas, as shown in Fig. 7. Some of
the hard X-rays are absorbed through ionization and re-emitted as
softer photons through recombination, bremsstrahlung, or line
emission.  For each model, we calculate $f_{\rm rep}$, the ratio of the
total luminosity emitted outward by the gas to the total input
luminosity.  For most models $f_{\rm rep} < 0.1$, and at most 50\% of
the reprocessed emission emerges in soft X-rays (0.1--2 keV).  There
are a few cases for the $\Delta R=0.1R$ shells that have $f_{\rm rep} >
0.2$, but in these only 10\%--20\% of the emission emerges in soft
X-rays, with most coming in the EUV.  In addition, the spectral shape
of the soft emission does not match observations. There is little
bremsstrahlung continuum, and the radiative recombination and atomic
lines produce spectra with discrete features and insufficient
flux at $\sim$1 keV.

To show this quantitatively, we used the output XSTAR spectra to
create table models in XSPEC and fitted these models to the observed
spectrum of \rxj.  We took data from the \asca\ GIS2 observation of
1993 May \citep{kohn00} from the HEASARC archive.  We used the
standard event-screening criteria and extracted the source and
background spectra from regions similar to those shown in Fig. 1 of
\citet{kohn00}.  We fitted the data in the range 0.7--7 keV.

For each XSTAR run, the model consisted of the input power law
spectrum plus the outward emitted spectrum, modified by typical
interstellar $N_{\rm H}=4\times10^{20}$ \cdens. We fitted these models to
the data, allowing the relative luminosities of the input and emitted
component to vary between 0.01 and 100 times that calculated by XSTAR
(indicating more or less reprocessing gas than in our model shells).
The models fitted, all having $\chi^2_{\rm \nu} \ge 2.5$ and most
having $\chi^2_{\rm \nu} > 7$.  In general, the reprocessed spectra have
discrete line features and too little flux at $\sim$1 keV.  In
addition, most spectra are underluminous;
for the model with the best fit ($Z=0.01$, $R=10^{12}$ cm, $N=10^{23}$ \cdens,
and $\Delta R=0.1 R$), the condition $\chi^2_{\rm \nu}=2.5$ requires an outward
luminosity 90 times that output by XSTAR.

To allow for reprocessing by gas at several different radii and
densities, we also performed spectral fits with multiple models
simultaneously.  For each $Z$ we included the emitted spectra from the
six models with the most luminous emission in soft X-rays (0.1--2 keV)
plus the input power law.  We allowed the normalization of each
emitted spectrum and the input power law to vary, and the results are
shown in Fig. 8.  Even when combining emission from several sets of
gas parameters, there remains a decrement at low energies, which is
not significantly affected by changing $N_{\rm H}$. As shown in \S\ 4,
this required soft flux could not simply be part of the emission from
the neutron star because of the high temperatures required.

Of course, the photoionized gas in XBP systems is far more complex
than we have modeled here, containing shocks, wakes, and other density
gradients, and in the case of Be stars, a circumstellar disk.  We have
not considered small regions of very dense gas ($n > 10^{12}$ \vdens) at
large radii, such as those present in simulations of SMC X-1
\citep{blon95} and suggested by the lines seen by \citet{vrti01}.  We
note, however, that such dense regions subtend a small angle to the
neutron star and so do not contribute significantly to the
reprocessed flux.  There also could be variations in the shape of the
photoionizing spectrum seen by different parcels of gas, which would
affect the photoionization emission in complex ways that would require
a much more detailed study to model completely.  Thus, it is in
principle possible that a particular configuration might produce the
soft spectrum seen in \rxj.  However, our analysis suggests that in
general photoionized emission is too weak and would have an
incorrect spectral shape.  In \rxj, as in the other luminous XBPs, the
soft excess most likely has a different origin.

\subsection{Higher optical depths}
Because XSTAR does not include Comptonization, our models have not
covered the case of Compton-thick ($N > 10^{24}$ \cdens) gas at radii
close to the neutron star where high densities may be present.
\citet{ross79} models this regime with code that explicitly calculates
the radiative transfer in Compton-thick gas.  The models show that for
$R \sim 10^8$ cm, high densities ($n \sim 10^{18}$ \vdens), and
$L_{\rm X}=2\times 10^{37}$ \ergs, there is still some decrement from
a blackbody shape at X-ray energies, even for high Thomson optical
depths $\tau_{\rm T} \sim 3$--5.  In this regime there are strong
emission features of oxygen and silicon at $\sim$0.8--1.2 keV, which
are not seen in the soft excess of most XBPs but may correspond to the
$\sim$1 keV emission lines found in Her X-1.  This suggests that close
into the neutron star, where we expect high gas densities, the gas must have high Thomson optical depths to produce the
blackbody-shaped spectrum of the soft excess.  We discuss this regime
in the next section.

\section{Reprocessing by optically thick material}
Reprocessing might occur in gas that is completely optically thick, with
$\tau_{\rm T} > 5$.  Such material is naturally expected to reside in
the accretion disk, for example, and when illuminated with hard X-rays
it re-radiates in a blackbody-like spectrum.  If the surface area
of this region is of the appropriate size, it can reproduce the
observed temperatures and luminosities.  The exact geometry of the
accreting material is unknown, but for simplicity we use the
phrase ``reprocessing by the inner disk'' to refer to all shapes
(shells, warped surfaces, etc.)  of optically thick gas at the
disk's inner edge.

To begin we consider a thick partial spherical shell centered on the
neutron star, subtending a solid angle $\Omega$ at the X-ray source
(see Fig. 9). Any photons that strike this gas are absorbed and
reprocessed, so that for isotropic emission,
$$L_{\rm soft}=\Omega L_{\rm X}.$$ If the radius of this shell is $R_{\rm BB}$, the
reprocessing region has an area
$$A=\Omega 4 \pi R_{\rm BB}^2.$$
For isotropic blackbody radiation,
$$L_{\rm soft}=\Omega 4\pi R_{\rm BB}^2 \sigma T_{\rm BB}^4.$$  Thus we have 
$$R_{\rm BB}^2 = \frac{L_{\rm X}}{4 \pi \sigma T_{\rm BB}^4}.$$ This
differs from some estimates for $R_{\rm BB}$, which assume all the
soft luminosity is radiated from a full sphere at $R_{\rm BB}$.  Our
formula is more appropriate if the blackbody comes from reprocessing
by a partially-covering, disk-like structure.  For typical $L_{\rm
X}=2\times10^{38}$ \ergs\ and $kT_{\rm BB}=0.17$ keV, we obtain $R_{\rm
bb}=10^{8}$ cm. This value is close to the typical radius of the
magnetosphere for the luminous XBPs, suggesting that the reprocessing
gas may be at the inner edge of the accretion disk.

\begin{figure}
\epsscale{1}
\centerline{\includegraphics[width=3.5in]{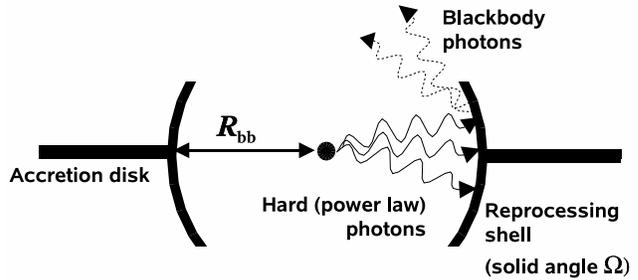}}
\caption{Schematic of the simple picture used to calculate the
blackbody emission radius $R_{\rm BB}$.}
\end{figure}

\subsection{Her X-1 and Cen X-3}
There is evidence for the above picture in Her X-1 and Cen
X-3. In Her X-1, the 35-day periodicity is attributed to the blocking
of the central neutron star by a precessing disk
\citep[e.g.,][]{gere76}.  Detailed geometric models for the disk
have been used to explain phenomena such as X-ray dips
\citep{shak99} and varying pulse profiles \citep{scot00}.  The
anomalous low states observed in 1983, 1993, and 1999 in which
the source remained in the low state for several 35-day periods, have
also convincingly been shown to be due to occultation by the disk
\citep{cobu00,manc03}.

Observations suggest that this blocking disk reprocesses the hard
X-rays to produce the soft excess. \citet{endo00} performed
pulse-phase resolved spectroscopy on \asca\ data taken during the
main-on state ($\phi_{35} \sim 0.1$).  As described in \S\ 2, they
found that the soft blackbody component and line emission features
pulsed 230\dgr\ out of phase from the narrow power-law pulses, an
offset seen earlier by \citet{mccr82}.  The authors proposed that the
tilted inner accretion disk absorbs the hard X-rays as the pulsar
points away from the line of sight, giving rise to the blackbody and
lines, which are emitted back toward the observer out of phase with
the hard pulses. For the spectral parameters found by \citet{endo00},
$R_{\rm BB} \sim 5 \times 10^7$ cm.  They showed that the cooling
timescale of the irradiated gas is a small fraction of a second, so it
is possible for the reprocessed soft X-rays to pulse along with the
hard component.

\citet{rams02} analyzed observations with \xmm\ taken at
$\phi_{35}=0.17$ (close to main-on), 0.26 (low state), and 0.60
(short-on state).  They examined how the phase shift between the soft
excess and hard component varies with \phip.  They found that the
profiles have phase offsets that vary from $150^{\circ}$ at $\phi_{35} =
0.17$ to $-10^{\circ}$ at $\phi_{35} = 0.26$ and
$90^{\circ}$ at $\phi_{35} = 0.6$.  They argued that this
continuous change in offset is expected from reprocessing by a tilted
precessing disk.

For Her X-1, the range of observed values $kT_{\rm BB} \sim0.1$--0.16 keV and
$L_{\rm X} \sim 2\times 10^{37}$ \ergs\ gives $R_{\rm BB} \sim
0.5$--$1.2\times 10^8$ cm. An estimate of the magnetospheric radius
\citep{fran02} is
$$R_{m}\sim0.5
R_{\rm A}\sim1.5\times10^{8}m_1^{1/7}R_6^{10/7}L_{37}^{-2/7}B_{12}^{4/7} {\rm
cm,}$$ where $R_{\rm A}$ is the standard \alfven\ radius.  Here $m_1$ is the
mass of the neutron star in $M_{\sun}$, $R_6$ is its radius in $10^6$
cm, $L_{37}$ is the X-ray luminosity in $10^{37}$ \ergs, and $B_{12}$
is the neutron star surface magnetic field in $10^{12}$ G.  For Her X-1 ($R_6 \sim
1$ and $B_{12} \sim 4$), $R_{m} \sim 3\times10^8$ cm.
\citet{scot00} calculate $R_{m}$ using several other models
\citep[i.e.][]{fing96,kira88,aly80} and find they all give $R_{\rm
M}\sim (3$--$4)\times10^8$ cm. Thus, $R_{m}$ is larger than $R_{\rm
bb}$ by as much as a factor of 4 or more.  We discuss possible explanations
for this discrepancy in the next subsection.

In Cen X-3, the soft excess is less well studied but is again
consistent with reprocessing by the disk.  \citet{burd00} found a
bright soft excess in the \bepposax\ spectrum, with $kT_{\rm BB}=0.11$
keV and containing 58\% of the total unabsorbed flux.  With $L_{\rm X}
\sim 10^{38}$ \ergs, we have $R_{\rm BB} \sim 2.4\times10^8$ cm, while
$B_{12}\sim3$ gives $R_{\rm M} \sim 1.5 \times 10^8$ cm.  As in Her
X-1, the blackbody and iron line are likely produced by reprocessing
at the magnetosphere.  \citet{kohm01} tested this picture by examining
aperiodic oscillations of the X-ray flux and measuring the time delay
of the iron lines with respect to the other bands.  They found a
systematic time delay of $0.39\pm1.0$ ms, which implies a light-travel
distance of $\sim$$1.7\times10^8$ cm.  This distance is comparable to
$R_{\rm BB}$ and $R_{m}$ and so supports the interpretation of disk
reprocessing.

\subsection{SMC X-1 and LMC X-4}
 Although SMC X-1 and LMC X-4 are more luminous than Her X-1, they
all show remarkable similarities including a sinusoidal soft pulse
profile, and a long-term periodicity, which is also attributed to a
blocking, precessing disk.  These similarities indicate that optically
thick reprocessing may be the source of the soft excess in these
systems as well.

Given the temperatures and luminosities for SMC X-1 and LMC X-4 from
\citet[see Table 1]{paul02} we find $R_{\rm BB} = 1.3$ and
$1.1\times10^8$ cm, respectively.  For LMC X-4 , $B_{12}\sim10$
indicates $R_{m}\sim2.9\times10^8$ cm.  This gives $R_{\rm
bb}/R_{m}=0.4$, compared with $R_{\rm BB}/R_{m}\sim0.2$--$0.4$
for Her X-1, depending on the value for $T_{\rm BB}$, and $\sim$1.5
for Cen X-3.  The values for $R_{\rm BB}$ and $R_{m}$ are shown
for comparison in Table 5.

In SMC X-1, no cyclotron feature has been detected, so there is no
magnetic field estimate. For SMC X-1 to have $R_{\rm BB}/R_{\rm
M}=0.4$, as in Her X-1 and LMC X-4, would require $B_{12} \sim 20$.
Such a high magnetic field would give a cyclotron line at $\sim$200
keV and so may be difficult to detect with current instruments.
However, \citet{maki99} suggest that the relatively low spectral
cutoff energy (6 keV) in SMC X-1 may indicate a smaller magnetic field
than Her X-1, so the question is still open. Another estimate for the
inner edge of the accretion disk is the corotation radius
$$R_{\rm cor}=(GM/\Omega_{\rm NS}^2)^{1/3},$$ which is the radius at
which the Keplerian angular velocity of the disk is equal to
$\Omega_{\rm NS}$, the angular velocity of the neutron star.  For SMC
X-1, $R_{\rm cor}=1.3\times10^8$ cm, which is equal to $R_{\rm BB}$.

\citet{paul02} argue that because of discrepancies between $R_{\rm
bb}$ and $R_{m}$, disk reprocessing may not be at work in the
luminous XBPs.  However, Her X-1 and Cen X-3 show substantial evidence
for reprocessing by the disk, yet have markedly different ratios
between $R_{\rm BB}$ and $R_{m}$. It is therefore difficult to
base conclusions on the exact values obtained for $R_{\rm BB}$ and
$R_{m}$.  This uncertainty exists for several reasons, some of
which are treated in detail by \citet{scot00}.  First, there may be
errors in our estimate of $R_{m}$, largely because the standard
models for the magnetospheric radius are incomplete.  The interaction
of the accretion disk with the magnetic field is poorly
understood, but certainly a high rate of mass flow from the disk onto
the magnetic field lines has some effect on the field
\citep[e.g.,][]{agap00,aly90}.  Whether this makes $R_{m}$
larger or smaller is uncertain.  It may also be that the inner edge of
the disk does not actually reside at the radius that we define as the
magnetosphere, but that some gas exists inside this radius.  Again,
such a process is not fully understood and warrants more
investigation.

\begin{deluxetable*}{lcccc}
\tabletypesize{\scriptsize}
\tablecaption{Blackbody and magnetospheric radii for selected XPBs}
\tablehead{
 \colhead{} & \colhead{Her X-1} &  \colhead{Cen X-3}
 & \colhead{LMC X-4} & \colhead{Vela X-1}}
\startdata
$L_{\rm X}$ (10$^{38}$ ergs s$^{-1}$) 	& 0.2 		& 1.0  	& 1.2 	& 0.01 \\
$B$ ($10^{12}$ G) 			& 4 	& 3.5  	& 10 	& 2.4 \\
$kT_{\rm BB}$ (keV) 			& 0.16, 0.1 	& 0.11 		& 0.17 		& 0.2 \\
$R_{\rm BB}$ (10$^8$ cm) 		& 0.49, 1.2 	& 2.3  	& 1.1 	& 0.07 \\
$R_{m}$ (10$^8$ cm) 		& 2.9 		& 1.5  	& 2.9 	& 4.5 \\
$R_{\rm BB}/R_{m}$ 			& 0.2, 0.4 	& 1.5 	& 0.4 	& 0.01
\enddata
\end{deluxetable*}

Even if the standard formulae are accurate and the accretion disk
really is truncated at $R_{m}$, there are uncertainties in the
parameters that are used to calculate $R_{m}$.  $B$ is taken from
measurements of a cyclotron feature that may arise not in the main
dipole field but in a region of stronger or weaker field near the
accretion flow.  Since $R_{m} \propto B^{4/7}$, a 25\% error in
the $B$ field strength would cause a $\sim$15\% error in $R_{m}$.
There is also uncertainty in $L_{\rm X}$ due to errors in the distance
to the source, uncertainty in the spectral parameters, and variation
of the source with time. Since $R_{m} \propto L_{\rm X}^{-2/7}$, a
25\% error in $L_{\rm X}$ would cause a $\sim$7\% error in $R_{\rm
M}$.  There are also uncertainties in the masses and radii of neutron
stars.  Masses of solitary pulsars seem to be tightly peaked around
1.4 $M_{\sun}$ \citep{thor99}, but for accreting pulsars the masses
may be greater.  For example, from radial velocities of the companion
to Vela X-1, \citet{quai03} obtain $M = 2.27\pm0.17 \ M_{\sun}$.
Since $R_{m} \propto M^{1/7}$, such a mass variation would
correspond to an increase in $R_{m}$ of $\sim$10\%.  Estimates of
$R_{\rm NS}$ vary from 0.7 to 1.6 $\times 10^6$ cm, which can
vary $R_{m}$ by another $\sim$20\%.  The overall effect of these
uncertainties is to create substantial uncertainty in $R_{m}$ which
may be as high as $\sim$30\% or greater.

There may also be errors in our estimate of $R_{\rm BB}$.  The
spectrum of the reprocessing material is almost certainly not a
single-temperature blackbody.  A hot disk coronona could
Compton-scatter some of the blackbody photons, giving an observed $T$
higher than the true blackbody temperature and causing an
underestimate of $R_{\rm BB}$. On the other hand, the reprocessing
surface may be tilted with respect to the hard X-ray flux, which would
increase the emitting area at a given radius and thus decrease $R_{\rm
bb}$.  Viewing angle effects due to non-isotropic emission could also
be important.

Given these uncertainties, the correspondence between $R_{m}$ and
$R_{\rm BB}$ to a factor of 2 or so suggests that reprocessing by the
inner disk can explain the spectra of LMC X-4 and SMC X-1.  Such a
model arises naturally from the geometry of these sources, and, unlike
the other processes we have considered, it reproduces the pulsations
and high luminosities of the soft component.

\subsection{\rxj\ and \xtej}
Transient Be/XBP systems are expected to form accretion disks
\citep{haya04}, so inner disk reprocessing might occur.  In \rxj\ and
\xtej, there is no magnetic field measurement, so estimates of $R_{\rm
M}$ are difficult.  A typical value of $B$ for XBPs ($B_{12}=4$)
gives $R_{m}\sim 1.5 \times10^8$ cm for \xtej.  This is of order
the $R_{\rm BB}\sim0.8\times10^8$ cm obtained from the \asca\ spectrum by 
\citet{yoko00}.  \rxj\ has a similar luminosity and spectrum, and so
may also have disk reprocessing.  As in SMC X-1 and LMC X-4, this
is the only mechanism that can produce the spectral shape and high
$L_{\rm soft}$ in these systems.

A puzzle is the lack of pulsations in the soft component of \rxj. One
explanation is that, unlike Her X-1, SMC X-1, and LMC X-4, the disk
may be at low inclination.  In this case, the whole reprocessing region
would present roughly the same angle to the observer, so we would see
a brightly emitting disk at all phases in the pulsation.  Another
possibility is that electron scattering from a large disk wind or
corona, of size $\sim$$10^{10}$ cm, could wash out the pulsations in
the soft component.

\subsection{Vela X-1 and \fouru}
In Vela X-1, reprocessing by the disk is {\it not} the main source of
the soft excess.  Given $L_{\rm X} \sim 10^{36}$ \ergs\ and $B_{12}
\sim 2.4$, we obtain $R_{m} \sim 5 \times 10^8$ cm.  Although a
blackbody fit has not been performed to the Vela X-1 soft component,
we note that \citet{paul02} were able to fit both blackbody and
bremsstrahlung models to SMC X-1 and LMC X-4 spectra, giving $kT_{\rm
bb}=0.17$ and 0.18 keV, and $kT_{\rm tb}=0.3$ and 0.5 keV,
respectively.  For Vela X-1, $kT_{\rm tb}=0.5$ keV, so we assume
an equivalent $kT_{\rm BB} \sim 0.2$ keV.  This gives $R_{\rm BB} \sim
7 \times 10^6$ cm, or $R_{\rm BB}/R_{m} \sim 0.01$, which is
extremely small.  In other words, Vela X-1 is not luminous enough to
produce a reprocessed blackbody soft component at $kT_{\rm BB}\sim0.2$
keV and over an area $\sim R_{m}^2$.  On the other hand, because of the
relatively low luminosity, the soft component is well explained by
emission from diffuse gas excited by collisions and photoionization,
as discussed in \S\S\ 5--6, respectively.

The situation in \fouru\ is difficult to evaluate because the distance
to the source, and thus $L_{\rm X}$, is poorly constrained.  The different
possibilities for a Roche lobe-filling companion to \fouru\ imply
different distances: 1 kpc for a 0.02 $M_{\sun}$ white dwarf, 3 kpc for
a 0.08 $M_{\rm \sun}$ partially degenerate, hydrogen-depleted star, and 36
kpc for a 0.6 $M_{\rm \sun}$ helium burning star \citep{chak98b}.
\citet{chak98b} found that optical photometry is consistent with emission coming mostly from an X-ray heated disk, which
implies a distance in the range of 5 kpc $\lesssim D \lesssim 13$ kpc.

\citet{orla98a} find that for a blackbody soft excess, the radius of
the emitting region is $1.5 \times D_{\rm kpc}$ km.  For $D \sim 5$
kpc and thus $L_{\rm X} \sim 2 \times 10^{36}$ \ergs, this region is
the size of the neutron star, so soft excess may be thermal emission
from the star's surface.  To evaluate the possibility of reprocessing
by the disk, we take $B_{12}=3.5$, $kT_{\rm BB}=0.3$ keV,
and $L_{\rm X}=7.7 \times 10^{34} D_{\rm kpc}^2$ and find that
$R_{\rm BB}$ is of order $R_{m} \sim 9 \times 10^7$ cm only at $D \sim
100$ kpc.  This is an unphysically large distance.  We conclude that
disk reprocessing is {\it not} at work in this system.

\subsection{Pulsations from disk reprocessing}
To conclude this section, we show that soft pulses can arise from disk
reprocessing in the luminous XBPs.  Following the calculation of
\citet{endo00}, the total thermal energy of the soft emitting region
is simply the volume of the region times its thermal energy
density
$$E_{\rm BB}=\Omega 4 \pi R_{\rm BB}^2 d \left ( \frac{3}{2} n kT_{\rm
bb} \right ),$$ where $d$ is the penetration depth of hard X-rays into
the heated region.  We assume that most X-rays penetrate no more than
a few Compton depths, so that $nd \sim 3/\sigma_{\rm T} \sim 10^{25}$
cm$^{-2}$.  For typical $R_{\rm BB} = 10^8$, $\Omega/4 \pi\sim 0.1$,
and $kT_{\rm BB} = 0.17$ K, we have $E_{\rm BB} \sim 10^{32}$ erg.
The cooling timescale is simply $t_{\rm cool}=E_{\rm BB}/L_{\rm
soft}$, which for $L_{\rm soft} \sim 10^{37}$ \ergs\ gives $t_{\rm
cool} \sim 10^{-5}$ s.  As in Her X-1, this is much shorter than the
pulsation period, so soft pulses should be possible for this
reprocessing mechanism.

\section{Discussion}
We have found that only a few of the models that have been proposed
can successfully account for the observed soft excess in XBPs. In
addition, the origin of the soft excess depends strongly on the
luminosity of the source.  For systems with high $L_{\rm X} \gtrsim
10^{38}$ \ergs\ (SMC X-1, LMC X-4, \rxj, \xtej, and Cen X-3), the soft
excess can be explained {\it only} by reprocessing of the hard X-rays
by optically thick accreting material.  However, for systems with
$L_{\rm X} \lesssim 10^{36}$ \ergs, the soft excess is emitted by
diffuse gas through collisional heating or photoionization
(Vela X-1, RX J0101.3--7211, and AX J0103--722), or is possibly thermal
emission from the neutron star's surface (\fouru, X Per).

The more luminous XBPs probably also emit soft X-rays from a diffuse
cloud and from the surface of the neutron star, but these components
are dominated by the flux from the power law and
from disk reprocessing.  On the other hand, in lower $L_{\rm X}$
systems we do not expect disk reprocessing to be important, since for
blackbody emission at $R_{\rm BB}=R_{m}$, we have $T_{\rm BB} \propto
L_{\rm X}^{1/4}/R_{m}^{1/2}$.  Since $R_{m} \propto L_{\rm
X}^{-2/7}$, this gives $T_{\rm BB} \propto L_{\rm X}^{11/28}$.  Thus for
lower $L_{\rm X}$ the temperature of the disk moves toward the EUV
and out of the soft X-ray band.

For XBPs with moderate $L_{\rm X} \sim 10^{37}$ \ergs, it
appears possible to have disk reprocessing (Her X-1), emission by
diffuse gas (\exo), or possibly both (\aofive).  We thus have
something of a continuum, in which the soft excess from more luminous XBPs
comes primarily from disk reprocessing, while in less luminous ones it
originates in optically thin emission or thermal emission from the
neutron star's surface.

In addition to these general conclusions, this work raises a number of
puzzling questions.  One is the apparently high brightness temperature of the
power-law component in SMC X-1 (\S\ 4.1).  This may suggest that the
accretion column is hotter or larger in extent than models predict.
For Vela X-1, the question remains whether the soft excess emission is
powered mainly by photoionization or collisional energy sources (\S\S\
5--6).  In \rxj, where disk reprocessing is likely at work, the
complete absence of soft pulsations (\S\ 7.3) is also worth further
investigation.

It is important to examine the lack of exact correlation between
$R_{\rm BB}$ and $R_{m}$ in the sources where we believe
reprocessing by the accretion disk is at work.  We have discussed a
number of uncertainties in these calculations in \S\ 7.2, but this
problem warrants further study.  For example, a thorough investigation
of non-blackbody effects in the disk reprocessing can help reduce
uncertainty in $R_{\rm BB}$.  Radiative transfer in an
obliquely illuminated model disk has been treated by
\citet{psal02}, and the structure and line emission of an illuminated
disk atmosphere has been modeled in detail \citep{jime01,jime02}.
Using such models to analyze the soft emission from XBPs can help
diagnose the gas conditions near the disk's inner edge.

There remains the question of how the accreting material, which is
expected to be in a geometrically thin disk, is able to subtend a
substantial solid angle to the central neutron star.  Interactions
with the magnetic field are likely to be key in this process.
For example, \citet{spru90} have shown analytically that accreting matter can be
tied to the magnetic field in equilibrium positions off the midplane
of the disk, near its inner edge.  Numerical, three-dimensional magnetohydrodynamic
simulations of accretion onto rotating magnetized stars
\citep{roma03, roma02} show structures including funnel flows and
dense ``rings'' that develop near $R_{m}$.  These may be the sites
of the X-ray reprocessing, and the soft excess emission may provide
constraints on the properties of such structures.

We can also use the soft excess to study warping and precession of the disk,
particularly in Her X-1, SMC X-1, and LMC X-4.  A warping instability
in a centrally-illuminated disk was found theoretically by
\citet{prin96} and has been explored in a number of studies
\citep[e.g.,][]{wije99, lai99, malo97} that also predict the disk's
precession.  As \citet{rams02} have done in the case of Her X-1, it
would be interesting to examine SMC X-1 and LMC X-4 and see how the
reprocessing region on the disk moves as the disk precesses.

\section{Summary}
In this paper we have reviewed a number of observations of the soft
excess in X-ray pulsarsand explored the origin of this
feature. We find the following:

\begin{enumerate}
\im The soft excess is a {\it very common}, if not ubiquitous feature
of emission from XBPs.  Observations of XBPs that have not shown a
soft excess are most often limited by low flux, high absorbing column,
or insufficient soft sensitivity.  \im For luminous ($L_{\rm X}
\gtrsim 10^{38}$ \ergs) sources, the soft components have high
luminosities and soft spectral shapes that can only be explained by
reprocessing of hard X-rays from the neutron star by optically thick,
accreting material, most likely near the inner edge of the accretion
disk.  \im For less luminous ($L_{\rm X} \lesssim 10^{36}$ \ergs)
sources, \ the soft excess is due to other processes, such as emission
by photoionized or collisionally heated diffuse gas, or thermal
emission from the surface of the neutron star.  In these sources the
soft component cannot come from reprocessing by the accretion disk.
\im XBPs of intermediate luminosity ($L_{\rm X} \sim 10^{37}$ \ergs)
show either or both of these types of emission.  \im Future
observations of soft excess will allow us to explore aspects of
XBPs such as the properties of circumstellar gas, interaction of
accreting matter and the magnetic field, and warping and precession of
the accretion disk.
\end{enumerate}

\acknowledgements We thank Rosanne DiStefano, Craig Heinke, Joseph Neilsen, John
Raymond, and Saku Vrtilek for helpful discussions, and the referee for
useful comments. This work was supported by NSF grant AST
0307433 and NASA grant NAG5-10780.  The research has made use of
NASA's Astrophysics Data System, and of data obtained from the High
Energy Astrophysics Science Archive Research Center (HEASARC),
provided by NASA's Goddard Space Flight Center.

\bibliographystyle{apj}

\end{document}